\documentclass[preprint]{aastex62} 
\usepackage{savesym}

\savesymbol{tablenum}
\usepackage[output-decimal-marker={.}, exponent-product=\cdot, per-mode=fraction]{siunitx}
\restoresymbol{SIX}{tablenum}

\usepackage{graphicx}
\graphicspath{{figures/}}

\usepackage{amsmath, amssymb, bm}
\usepackage{tabularx}
\usepackage{multirow}

\DeclareSIUnit[]\rsun
{\text{R\ensuremath{_{\textup{sun}}}}}

\shorttitle{MHD simulations on the large-scale propagation of high-speed solar wind streams}
\shortauthors{Hofmeister et al.}

\begin{document}

\title{MHD simulations on the large-scale propagation of high-speed solar wind streams}
    \author[0000-0001-7662-1960]{Stefan J. Hofmeister}
    \affiliation{Columbia Astrophysics Laboratory, Columbia University, 538 West 120th Street, New York, NY 10027}

\begin{abstract}
We investigate the propagation of high-speed solar wind streams from their origin near the Sun to 1 AU using three-dimensional magnetohydrodynamic simulations. By tracking both global stream structure and individual plasma parcels, we assess how local in-situ measurements relate to the underlying plasma evolution. We find that high-speed streams are not parcel-preserving structures: commonly used diagnostics such as peak velocity, density, or temperature do not trace fixed plasma elements, and feature-based radial trends can therefore misrepresent the true evolution. Instead, velocity-based relationships provide a more robust framework for linking plasma parcels across heliocentric distances. Stream evolution is dominated by interaction regions, where compression leads to deceleration of fast wind, acceleration of slow wind, and significant heating. A boundary layer forms close to the Sun and can dominate narrow streams, biasing in-situ measurements toward lower apparent velocities. We show that three-dimensional transport, in particular latitudinal flows, redistributes mass and magnetic flux and reduces center-to-flank contrasts. While radial magnetic flux is conserved, the total field strength is not in spherical sampling geometries due to non-radial components. Finally, observed stream properties and geoeffectivity depend strongly on sampling location, stream geometry, and latitudinal magnetic deflection, introducing systematic variability and asymmetries in geomagnetic response.
\end{abstract}

\section{Introduction}

This paper follows the evolution of a high-speed solar wind stream from its origin at the Sun to its impact at Earth. High-speed streams form the fast component of the interplanetary solar wind. They originate in solar coronal holes, propagate through interplanetary space, and frequently drive geomagnetic disturbances when they interact with Earth’s magnetosphere \citep{cranmer2002, tsurutani2006}.

Within solar coronal holes, plasma is accelerated along magnetic field lines that are open to interplanetary space \citep{krieger1973, zirker1977} . Closed loops are mainly confined to low and intermediate coronal heights \citep{wiegelmann2004}, so that their absence at larger heights leads to an excess pressure in open field regions relative to their surroundings. This drives a lateral expansion of open magnetic field regions with height until the transverse magnetic pressure component reaches equilibrium, leading to a quasi-uniform coronal absolute magnetic field strength at large height \citep{altschuler1969, wiegelmann2017}. As the coronal plasma is effectively frozen into the magnetic field, the plasma outflow in coronal holes drags the magnetic field lines outward, producing an approximately radial configuration at large heights \citep{altschuler1969}. Coronagraphic observations and magnetic field models indicate that the transition from the highly-structured low corona to this radial configuration is largely complete by $2.5$~solar radii \citep{altschuler1977,levine1977}. The outflowing plasma then follows these radial field lines to even larger heights and eventually propagates radially away from the rotating Sun \citep{parker1965}.

Farther from the Sun, starting from the Alfvenic point which is estimated to be located at about $8$~solar radii, the dynamic pressure of high-speed streams exceeds both the magnetic and thermal pressure \citep{weber1967, kasper2021} . The bulk momentum of the high-speed stream starts to govern its propagation characteristics so that it continues propagating radially away from the Sun, dragging the magnetic field with it. Since the magnetic field embedded in the outflowing plasma also remains anchored in coronal holes at the rotating Sun, the  magnetic field of high-speed streams is wound  by the solar rotation into the well-known Parker spiral \citep{parker1958}. 

At the leading edge of a high-speed stream, the fast plasma overtakes the preceding slow solar wind, forming a compressed stream interaction region \citep{gosling1978}. When such a region driven by a high-speed stream reaches Earth, their interplanetary magnetic fields can reconnect with the magnetosphere, enabling solar wind plasma to enter magnetospheric current systems and drive geomagnetic disturbances \citep{gonzalez1987, gonzalez1994}. The strength of these disturbances varies systematically over the year and depends on the relative orientation of the interplanetary magnetic field and Earth’s dipole axis . Because the magnetic field in high-speed streams follows the Parker spiral, while Earth’s dipole axis changes its orientation relative to the Sun over the year, there are two periods—around March and September—when the conditions are particularly favorable for reconnection. This seasonal modulation, known as the Russell–McPherron effect, increases the likelihood that high-speed streams trigger geomagnetic activity during these times \citep{russell1973}.

Although high-speed streams are regularly studied since the 1970s and their general propagation behavior seems to be well understood, many aspects are also still unclear. In particular, there are two major open issues to be resolved: (1) high-speed stream acceleration, which is one of the major open questions in solar physics; and (2) predicting their detailed propagation and arrival properties at Earth, which is important for space weather research. Several recent spacecraft missions, most notably Parker Solar Probe \citep{fox2016} and Solar Orbiter \citep{muller2020}, aim to address these questions by providing measurements closer to the Sun than ever before and enabling multi-spacecraft studies for the evolution of solar wind structures through interplanetary space.

In this work, we focus on the propagation of high-speed streams through interplanetary space and their arrival properties at Earth. Such propagation is typically studied by comparing in-situ plasma measurements from two approximately radially aligned spacecraft at different heliocentric distances \citep{perrone2019,allen2021,perrone2022,henderson2025,damicis2026}  . Ideally, one would track the evolution of individual plasma parcels as they propagate outward from the Sun. In practice, however, spacecraft are never perfectly aligned in latitude and longitude. Although they may sample similar regions of a stream at different radial distances, they cannot reliably follow the same plasma parcels. Moreover, the relative position of each spacecraft within the stream generally remains uncertain, making it difficult to relate the measured properties to the global stream structure. Consequently, studies typically focus on readily identifiable quantities, such as the evolution of the locally measured peak velocity and the density within the velocity plateau around that peak. However, these quantities cannot be directly related to the evolution of individual plasma parcels or to the global structure of the stream.

To address these limitations, we use magnetohydrodynamic (MHD) simulations to investigate the properties, evolution, and geoeffectiveness of a high-speed stream from the Sun to Earth. In contrast to observations, MHD simulations allow us to track the evolution of individual plasma parcels while simultaneously retaining knowledge of the global stream properties. We initialize the simulations by prescribing the cross-sectional properties of a high-speed stream close to the Sun, propagate the stream to Earth, and analyze the evolution of both local plasma parcels and the global stream. We further assess the resulting geoeffectiveness as a function of stream size and the location of in-situ measurements within the stream. This approach enables us to place local measurements into their global context. By that, it provides solar wind data analysts and space weather researchers with a comprehensive view of high-speed stream propagation and clarifies several physical relationships that are often used—sometimes incorrectly—in the interpretation of solar wind observations.

The remainder of this paper is structured as follows. Section~\ref{sec:setup} describes the simulation setup. Section~\ref{sec:closeSun} examines the formation of the transversal high-speed stream profile close to the Sun. Section~\ref{sec:propagation} analyzes the radial evolution of stream properties during propagation from the Sun to Earth. Section~\ref{sec:si} focuses on the interaction between the high-speed stream and the preceding slow solar wind. Section~\ref{sec:1au} investigates the properties of high-speed streams and their geoeffectiveness at Earth dependent on the conditions near the Sun. Finally, Section~\ref{sec:summary} summarizes our conclusions.

\section{Setup of the simulations}  \label{sec:setup}

\begin{figure}[t!]
 \includegraphics[width=\textwidth]{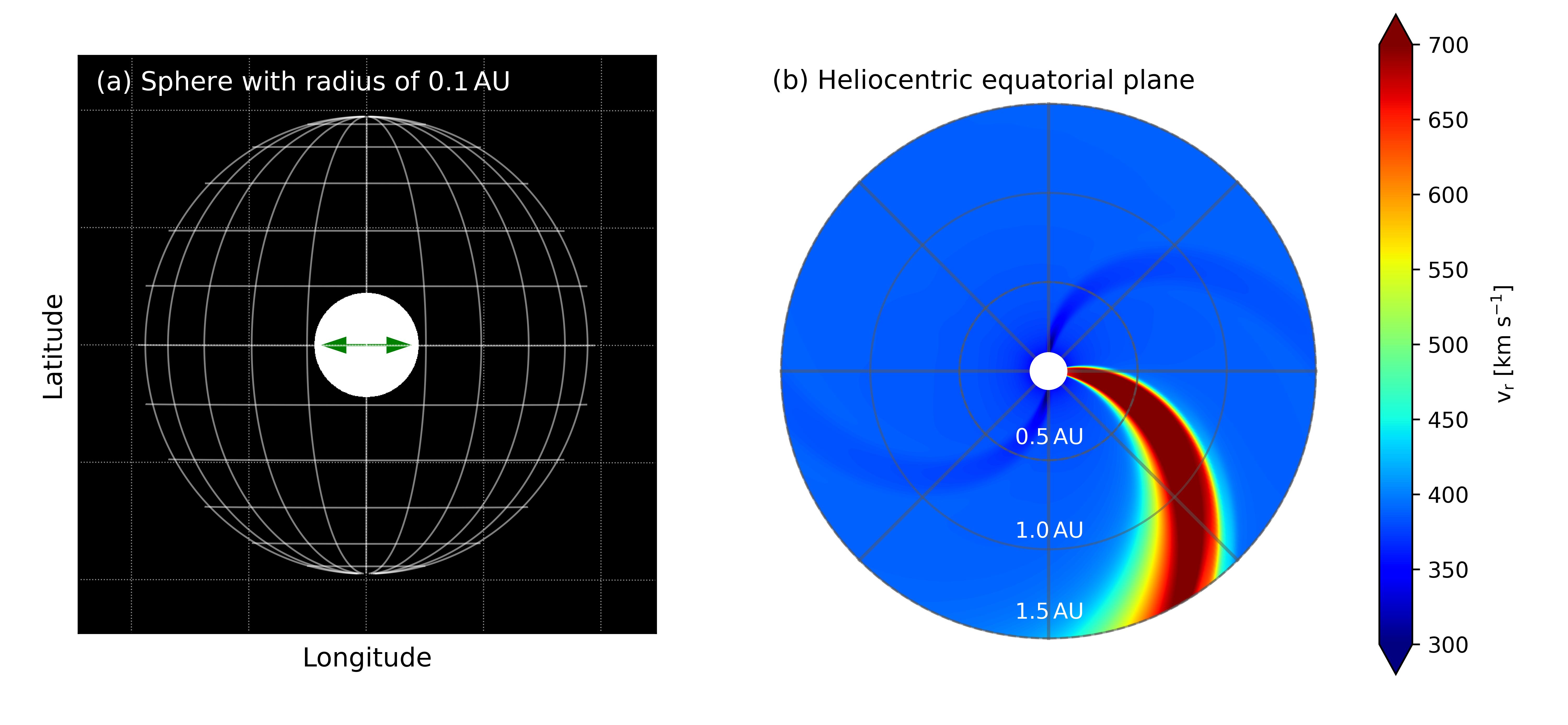}
\caption{Setup of the simulations. (a) At the inner boundary being located at \SI{0.1}{AU} from the Sun, we prescribe the footprint of a high-speed stream of variable size (white). (b) The resulting high-speed stream propagates through the heliosphere using MHD simulations, as illustrated in this snapshot of the heliocentric equatorial plane. }
\label{fig:setup}
\end{figure}

\begin{figure}[t!]
\centering
 \includegraphics[width=\textwidth]{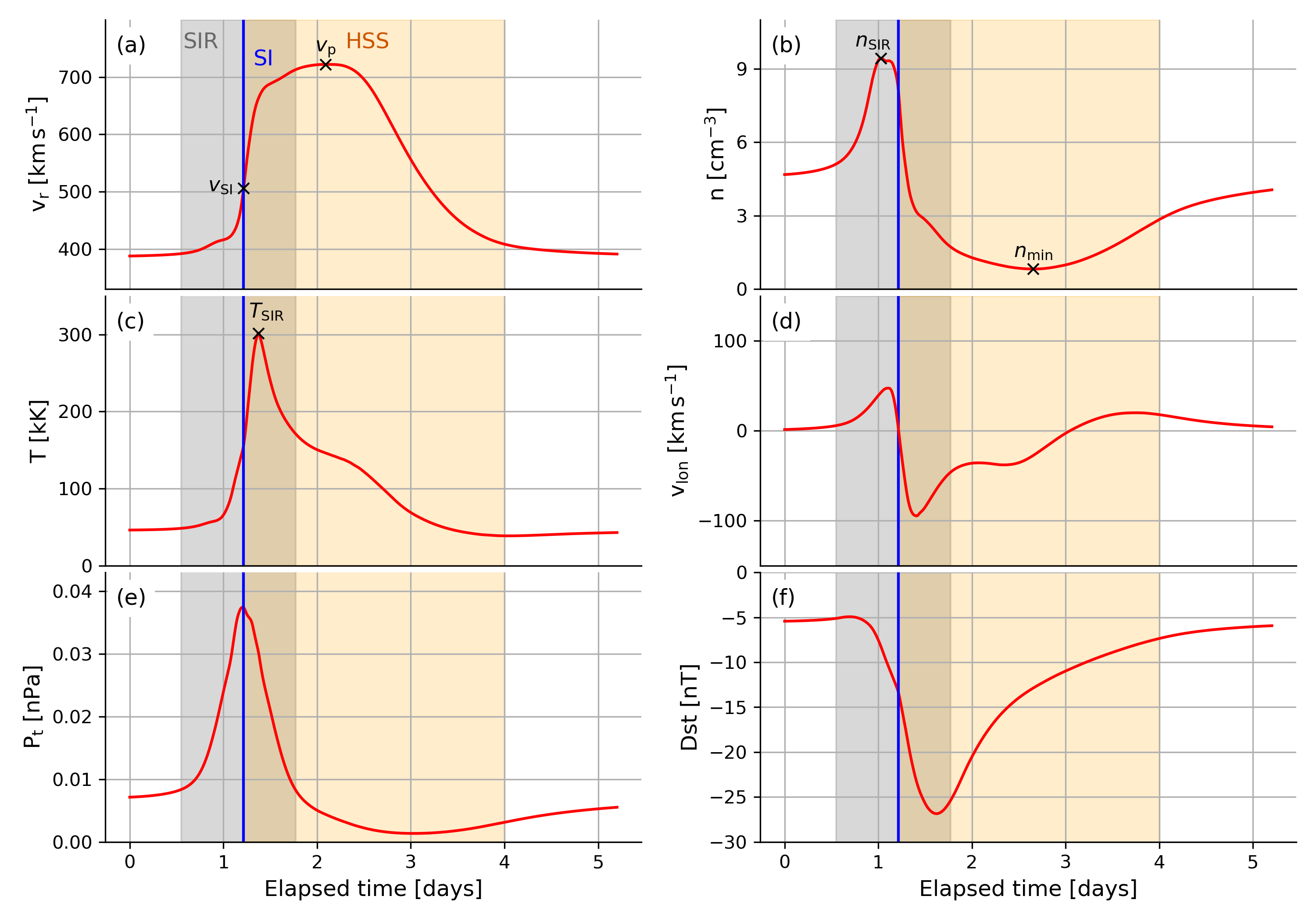}
\caption{Simulated high-speed stream and its geomagnetic response. (a) Velocity, (b) density, (c) temperature, (d) longitudinal velocity component, (e) total transverse pressure, and (f)~resulting Dst index versus the elapsed time in our simulations. The high-speed stream (HSS) is depicted by the light-orange shaded region. Light-orange shading indicates the high-speed stream (HSS). The gray shaded region marks the stream interaction region (SIR), which partially overlaps with the HSS. The stream interface (SI) separating the HSS from the preceding slow solar wind is shown in blue. }
\label{fig:hss_definition}
\end{figure}

We simulate the evolution of a high-speed solar wind stream with the European Heliospheric Forecasting Information Asset \citep[EUHFORIA;][]{pomoell2018}. EUHFORIA is comprised of a coronal model coupled to a three-dimensional, time-dependent MHD model of the inner heliosphere. In the present study, we restrict EUHFORIA to its heliospheric component and prescribe the solar wind directly at the inner heliospheric boundary, defined as a spherical surface with a radius of \SI{0.1}{AU} centered on the Sun.  This approach avoids uncertainties associated with the not-yet-understood acceleration of the solar wind below this height and allows us to construct a controlled, idealized heliospheric configuration.

To generate a high-speed stream, we prescribe at the inner boundary conditions at the solar equator a circular region of fast solar wind, which is embedded within a slow solar wind background. This is illustrated in Figure~\ref{fig:setup}(a). Each the fast and the slow solar wind are assigned uniform plasma properties, separated by a step-function transition. By systematically varying the diameter of the high-speed stream from a small diameter of \SI{3}{\degree} to large diameter of \SI{36}{\degree}, while keeping all other parameters fixed, we investigate how its cross-sectional extent near the Sun influences its evolution through interplanetary space and its signatures at \SI{1}{AU}.

At the inner boundary, we assign the fast stream a radial velocity of \SI{650}{km.s^{-1}}, a density of \SI{150}{cm^{-3}}, and a gas pressure of \SI{3.3}{nPa}, while the surrounding slow solar wind is prescribed with \SI{350}{km.s^{-1}}, \SI{500}{cm^{-3}}, and the same gas pressure. These values produce typical conditions at \SI{1}{AU} of roughly \SI{730}{km.s^{-1}} and \SI{1.5}{cm^{-3}} within the high-speed stream, and \SI{390}{km.s^{-1}} and \SI{5}{cm^{-3}} in the ambient slow solar wind. To reflect the equilibrium reached after the lateral expansion of coronal open magnetic field regions into the heliosphere, we initialize the fast and slow solar wind regions in lateral total pressure balance. We therefore impose a spatially uniform radial absolute magnetic field strength of \SI{325}{nT}, corresponding to a magnetic field strength \SI{1.5}{G} at the solar surface. This magnetic field setting results in a total heliospheric open flux of \SI{9e14}{Wb}, which approximates the conditions in the declining phase of Solar Cycles~$23$ and~$24$ \citep{frost2022}. To ensure global magnetic flux conservation, we include a heliospheric current sheet with a polarity inversion line given by $\varphi = \SI{-30}{\degree} \cos{\lambda}$, where $\varphi$ is the heliospheric latitude and $\lambda$ is the longitude, assigning opposite radial polarities to the northern and southern hemispheres. This current sheet serves only to balance the magnetic flux and has no significant impact on the high-speed stream dynamics, so it is neglected in the subsequent analysis.

We advance these initial conditions on a spherical grid extending from \SI{0.1}{AU} to \SI{1.5}{AU}, with a radial resolution of \SI{0.0055}{AU} and an angular resolution of \SI{1.0}{\degree}. Following an initial relaxation phase of $7$~days, the system reaches a quasi-steady, co-rotating state, which is presented in Figure~\ref{fig:setup}(b). We then sample the resulting solution with virtual spacecraft positioned at a fixed heliospheric longitude of \SI{0}{\degree}, while varying their heliospheric latitudes from \SIrange{0}{12}{\degree} and radial distances from \SIrange{0.1}{1.5}{AU}.

From the virtual spacecrafts, we extract the solar wind diagnostics, which is illustrated in Figure~\ref{fig:hss_definition}. The solar wind diagnostics comprises the three-dimensional solar wind velocity~$\mathbf{v}$, plasma particle density~$n$, temperature~$T$, magnetic field strength~$\mathbf{B}$, particle flux~$n\,\mathbf{v}$, and dynamic pressure~$0.5\,n\,v^2$. In addition, we investigate the evolution of the peak velocity $v_\text{p}$ and the minimum density $n_\text{min}$ of the high-speed stream, the propagation velocity of the stream interface between the high-speed stream and the preceding slow solar wind $v_\text{SI}$, and the peak plasma density $n_\text{SIR}$, peak temperature $T_\text{SIR}$, and peak magnetic field strength $\mathbf{B}_\text{SIR}$ within the compressed stream interaction region. Thereby, we determine the location of the stream interface from the zero-crossing of the longitudinal deflection angle of the velocity as illustrated in Figure~\ref{fig:hss_definition}(d), following \citet{gosling1978}. This zero-crossing marks the transition from westward-deflected slow solar wind plasma ahead of the stream interface to eastward-deflected high-speed stream plasma behind it. For visualization, we also show in Figure~\ref{fig:hss_definition} the location and extent of the compressed stream interaction region, defined by an enhancement in lateral pressure, i.e., the sum of gas and magnetic pressure, following \citet{richardson2018}. However, the precise spatial extent of the stream interaction region is of secondary importance for our study, as we determine the peak density and magnetic field strength within the compressed stream interaction region directly from the virtual spacecraft measurements, taking the global maxima of the density and magnetic field, respectively.

We estimate the geomagnetic Dst index by introducing a virtual Earth into our simulations. The Dst index quantifies the strength of the magnetospheric ring current in response to solar wind driving and is physically measured from variations in the horizontal magnetic field at four low-latitude ground stations distributed around the globe. Empirically, the response of the Dst index to solar wind conditions can be described by the differential equation of \citet{obrien2000}, in which the temporal evolution of Dst is governed by a balance between solar wind energy input and the decay of the ring current. The evolution of the ring current and the corresponding Dst index is described by
\[
\frac{d\text{Dst}^\star}{dt} = a\, Q - \frac{\text{Dst}^\star}{\tau}, \label{eq:dst}
\]
where $\text{Dst}^\star$ is the raw Dst index, representing the contribution of the ring current to the geomagnetic field before any correction for magnetopause compression. The term $Q$ describes the solar wind injection, given by
\[
Q = 
\begin{cases} 
v\,B_z + 0.49~\mathrm{mV/m}, & \text{for } v\,B_z < -0.49~\mathrm{mV/m}, \\[1mm]
0, & \text{for } v\,B_z \ge -0.49~\mathrm{mV/m},
\end{cases}
\]
where $B_z$ is the southward component of the solar wind magnetic field in geocentric solar magnetospheric (GSM) coordinates, i.e., along Earth's magnetic field axis. The coupling coefficient $a = 4.4~\mathrm{nT/(h\cdot(mV/m))}$ converts this driving into a rate of change of the ring current. The decay of the ring current is governed by the decay time
\[
\tau = 2.40 \,\exp\!\left(\frac{9.74}{4.69 + v\,B_z / \mathrm{mV/m}}\right)\,\mathrm{h},
\]
which depends on the strength of the driving solar wind, with typical values around 8 hours.
The observed Dst index measured on the ground is obtained by correcting $\text{Dst}^\star$ for magnetopause compression and other residual effects:
\[
\text{Dst} = \text{Dst}^\star + b\,\sqrt{0.5\, n\, v^2} - c,
\]
where $b = 7.26~\mathrm{nT/(nPa)^{-0.5}}$ accounts for the pressure correction due to magnetopause compression from the solar wind, and $c = 11~\mathrm{nT}$ represents residual offset effects. In a modern context, these equations can also be interpreted in terms of dayside and nightside reconnection, where solar wind driving enhances the ring current and magnetospheric relaxation governs its decay \citep{vasyliunas2006}.

MHD simulations generally tend to underestimate plasma compression in stream interaction regions. Consequently, the peak density and magnetic field strength in these regions, and thus the strength of the derived Dst index, are underestimated in our study. However, the results presented in the following sections are qualitatively robust and provide reasonable quantitative guidance, which is the main goal of our study.

\section{Properties of HSSs close to the Sun} \label{sec:closeSun}

\begin{figure}[t!]
 \includegraphics[width=\textwidth]{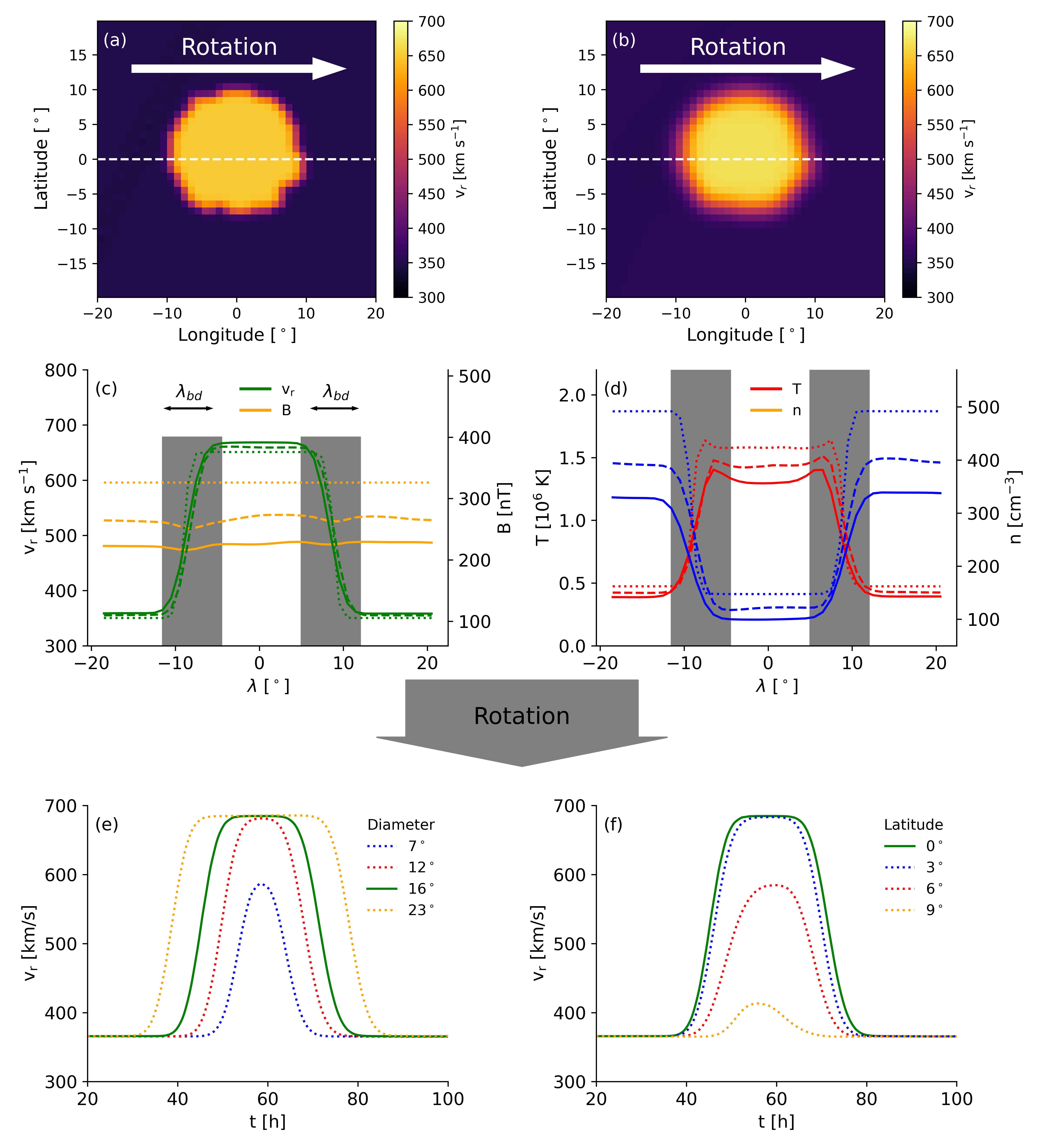}
 \caption{High-speed stream properties close to the Sun. (a) and (b) Solar wind velocity across the high-speed stream cross-section at heights of \SI{0.10}{AU} and \SI{0.12}{AU}, respectively. The latitudinal center of the high-speed stream is marked by the white dashed line. The direction of the solar rotation and the location of a virtual spacecraft (VSC) are annotated. (c) and (d) Solar wind radial velocity, density, temperature, and magnetic field strength along the slice through the latitudinal center of the high-speed stream. Solid lines show the properties at \SI{0.10}{AU}, dashed lines at \SI{0.11}{AU}, and dotted lines at \SI{0.12}{AU}. The size of the boundary region, $\lambda_\text{bd}$, is marked by the gray-shaded regions. (e) Solar wind velocities as recorded by a virtual spacecraft passing through the latitudinal center of the high-speed stream, as the high-speed stream passes beneath it with the solar rotation rate, for high-speed stream diameters of $7$, $12$, $16$, and~\SI{23}{\degree}. (f) Same as~(e), but for a fixed high-speed stream diameter of \SI{16}{\degree} while the virtual spacecraft passes at $0$, $3$, $6$, and \SI{9}{degree} latitude relative to the high-speed stream's latitudinal center.}
\label{fig:closetoSun}
\end{figure}

We initialize our simulations with a circular high-speed stream at the heliospheric equator, embedded in a slow solar wind environment and separated by a step-function transition, and propagate it through interplanetary space. Figure~\ref{fig:closetoSun}(a) and (b) show the velocity distributions across the high-speed stream cross-section and the surrounding slow solar wind at \SI{0.1}{AU} and \SI{0.12}{AU}, respectively, i.e., immediately after the solar wind enters the simulation domain. Already between the ghost cells, where we impose the boundary condition, and the inner boundary at \SI{0.1}{AU}, the discontinuity between the two regimes begins to smooth out. By \SI{0.12}{AU}, a stable boundary layer has developed between the high-speed stream and the surrounding slow solar wind. We further illustrate the evolution of the velocity, temperature, density, and magnetic field across this boundary layer along an equatorial cut through the stream, as shown in Figure~\ref{fig:closetoSun}(c) and (d). Across the boundary layer, the plasma properties transition smoothly between the high-speed stream and slow solar wind regimes, while the magnetic field varies only weakly due to the identical initial magnetic field strength on both sides. The width of the boundary layer stabilizes at approximately $\lambda_\text{bd} = \SI{7}{\degree}$, which is notably close to values inferred from in-situ measurements by ACE at L1 and by Ulysses during its fast latitude scan, reporting widths of about \SI{6}{\degree}.

We then place a virtual spacecraft at \SI{0.12}{AU} aligned with the latitudinal center of the high-speed stream and allow the stream, which co-rotates with the Sun, to pass beneath it. The spacecraft samples the temporal evolution of the solar wind plasma properties as the stream passes and continues to propagate radially into interplanetary space. Figure~\ref{fig:closetoSun}(e) shows the resulting temporal velocity profile as a function of the diameter of the high-speed stream. For large streams, a velocity plateau with a speed of about \SI{650}{km\,s^{-1}} is present, and its duration scales with the diameter of the stream. For small streams, however, no clear velocity plateau is observed, and the peak velocity remains well below the initialized value of \SI{650}{km\,s^{-1}}. This indicates that boundary effects influence the entire structure of small high-speed streams and reduce the flow speed in the stream center.

Next, we fix the diameter of the high-speed stream to \SI{16}{\degree} and examine the observed velocities as a function of the spacecraft's latitude relative to the center of the stream. The resulting velocity distributions are shown in Figure~\ref{fig:closetoSun}(f). When the spacecraft passes near the latitudinal center, it samples the core of the high-speed stream and recovers the peak velocity; when it passes near the flank, it primarily samples the lower-velocity plasma within the boundary layer. Since the boundary layer has a width of approximately \SI{7}{\degree} at both the northern and southern flanks of the stream, the boundary layer occupies a large area of the high-speed stream, making it likely that a real spacecraft would only measure the plasma properties in the flanks and not in the center of the stream. 

These results have several important implications for real-world observations. First, the formation of the solar wind boundary layer arises from dynamic propagation effects, as indicated by the smoothing of the initially step-function-like boundary with height. Consequently, observed variations of solar wind properties across high-speed stream boundaries do not necessarily reflect variations in solar wind acceleration at the edges of the source coronal hole, but can also result from the dynamical evolution during propagation.  
Second, when a spacecraft, such as Parker Solar Probe, measures lower peak velocities for small-diameter high-speed streams close to the Sun, this does not automatically imply lower initial acceleration in the solar atmosphere due to local physical conditions. Instead, such reductions can arise from boundary effects that slow the flow during the early propagation of the stream.  
Third, since the latitudinal position of real spacecraft relative to the stream center is typically unknown, it is often unclear whether the satellites sample the stream core or its flanks. Because plasma properties differ significantly between these regions, the measurements may not be representative of the entire stream. In real world data analysis, this effect of unknown satellite location relative to the stream center can introduce substantial scatter in empirical relationships, particularly between the by the satellites measured peak velocities of high-speed streams and the properties of their source coronal holes.

\section{Propagation of HSSs in interplanetary space} \label{sec:propagation}

Next, we examine the evolution of high-speed stream properties during their propagation through interplanetary space. Our analysis focuses on the radial evolution of the high-speed stream along the latitudinal center of a small high-speed stream with a diameter of \SI{10}{\degree} at the inner boundary, unless noted otherwise. We select this small-diameter stream because it illustrates the evolution of the plasma properties more clearly, although the qualitative behavior described below also applies to larger streams and to the properties in the flanks. The primary difference in the plasma evolution in the flanks arises from the distinct initial conditions near the Sun, where velocities and temperatures are lower and densities are higher compared to the stream center.

\subsection{Velocity, density, and temperature}\label{sec:vnt}

\begin{figure}[t!]
 \centering
 \includegraphics[width=.8\textwidth]{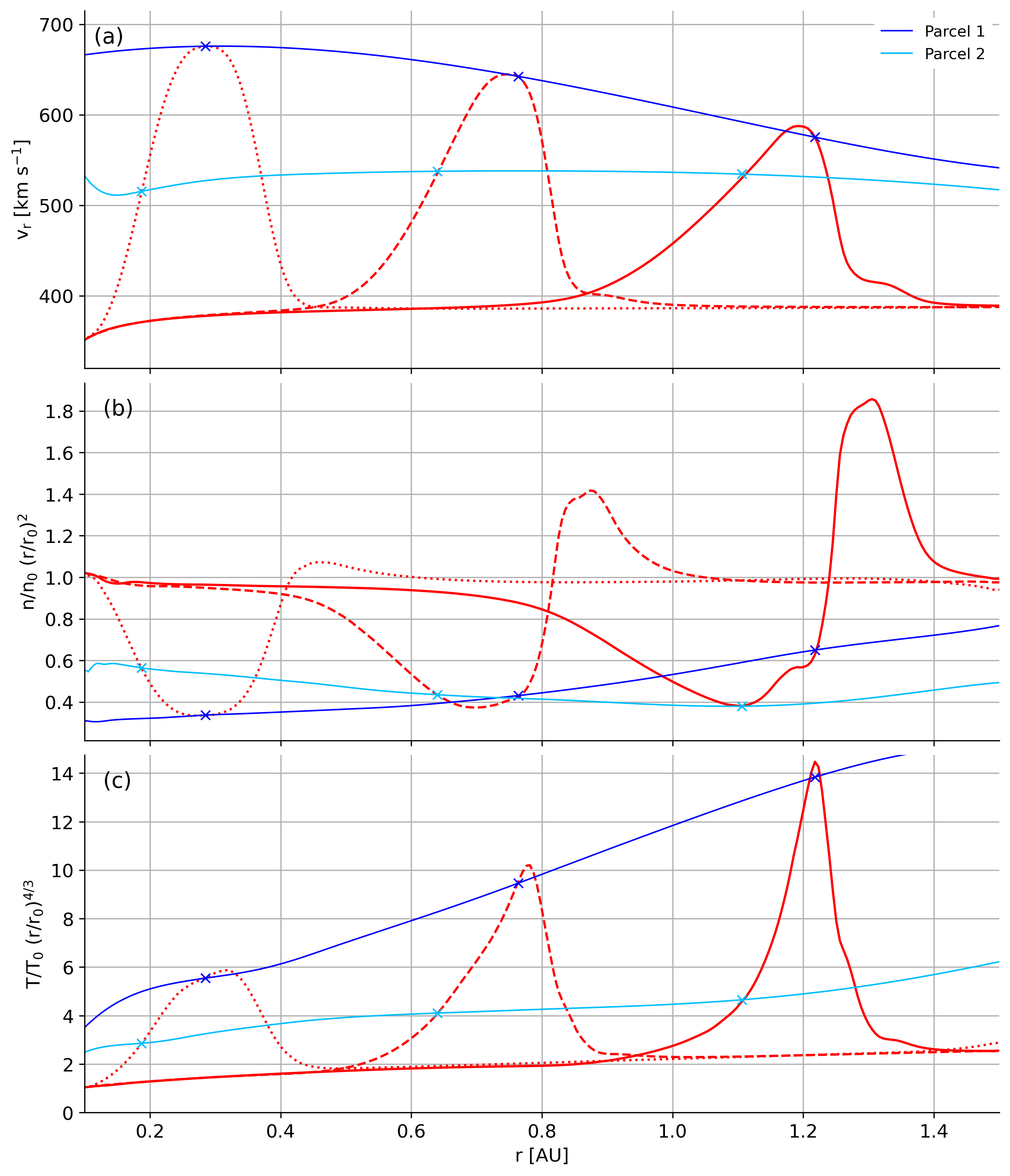}
\caption{Evolution of high-speed stream properties at the stream center. Panels show (a)~velocity, (b)~density, and (c)~temperature along a fixed radial direction after $1$~(dotted), $2.5$~(dashed), and $4$~days. in the simulations. The density and temperature have been scaled by $(r/r_0)^2$ and $(r/r_0)^{4/3}$, respectively, to remove the expected effects of spherical expansion. The scaled quantities are further normalized by the ambient slow solar wind density $n_0$ and temperature $T_0$ at the inner boundary, expressing the stream properties relative to the near-Sun slow solar wind. The blue and cyan curves trace the evolution of two selected plasma parcels within the stream; crosses mark their positions after $1$, $2.5$, and $4$~days.}
\label{fig:radial_vnt}
\end{figure}

We begin by examining the evolution of the high-speed stream velocity profile. Figure~\ref{fig:radial_vnt}(a) shows snapshots of the velocity along a constant radial direction at $1$, $2$.5, and $4$~days into the simulations. At $1$~day, i.e., close to the Sun, the velocity profile is symmetric and closely resembles the cross-sectional velocity distribution across the high-speed stream, as shown in Figure~\ref{fig:closetoSun}(e). By $2.5$~days, the velocity distribution has steepened slightly at the leading edge of the high-speed stream, while it flattens its tail. After $4$~days, the velocity profile exhibits a pronounced asymmetry, with a sharp rise at the front and an extended tail. Simultaneously, the peak velocity has decreased substantially from $680$~to \SI{590}{km\,s^{-1}}.  

To understand the origin of this velocity reduction, we tracked two plasma parcels in the simulations: Parcel~1 launched from the core of the high-speed stream, and Parcels~2 launched from the center of the trailing boundary region. Parcel~1 undergoes strong deceleration. This deceleration is caused by the stream interaction region, as illustrated in Figure~\ref{fig:hss_definition}, where compressed high-speed stream and slow solar wind plasma generates a pressure gradient that slows the incoming high-speed stream plasma while simultaneously accelerating the preceding slow solar wind. This interaction reduces the peak velocity of the high-speed stream and steepens the velocity profile at the front. In contrast, Parcel~2, launched from the high-speed stream tail, propagates with an almost constant velocity, demonstrating that not the entire high-speed stream decelerates. Beyond the stream interaction region, the pressure gradient is weak, allowing plasma parcels to propagate nearly unimpeded through interplanetary space. Since the velocity varies across the high-speed stream tail, the tail gradually disperses with radial distance, producing the observed flattening of the velocity profile in the tail with increasing distance to the Sun.

The density evolution, shown in Figure~\ref{fig:radial_vnt}(b), reflects the underlying physics of the velocity evolution described above. To isolate these effects, we remove the geometric $r^2$ density decrease due to spherical expansion and normalize the resulting density to the ambient slow solar wind density at the inner boundary. This representation highlights radial plasma compression and depletion relative to the ambient solar wind.

In this normalized representation, all prominent regions within the high-speed stream exhibit an increase in density with distance from the Sun. At the front, the peak density in the stream interaction region rises from $1.07$ to $1.86$, corresponding to an increase of \SI{74}{\percent}. Within the stream core, evaluated at the location of the peak velocity, the density increases from $0.34$ to $0.57$, i.e., by \SI{68}{\percent}. Even the minimum density in the tail shows a slight increase from $0.33$ to $0.38$, corresponding to \SI{13}{\percent}.  The increase in normalized density implies that the actual density decreases more slowly than expected from spherical expansion alone.

To relate these trends to the physical density evolution, we interpret them in terms of a power-law dependence of the form $n = n_0 \left(r/r_0\right)^\alpha$ for the underlying (non-normalized) density. Quantitatively, we obtain $\alpha = -1.48$ for the peak density in the stream interaction region, $\alpha = -1.63$ at the location of the peak velocity, and $\alpha = -1.91$ for the minimum density. All values are less negative than $-2$, confirming that the density reduces slower expected from spherical expansion alone.

However, these trends do not directly represent the density evolution of individual plasma parcels. By tracking the two selected parcels from above, we find that Parcel~1, launched from the high-speed stream core, increases in density from $0.34$ to $0.66$, i.e., by \SI{94}{\percent}, corresponding to $\alpha = -1.44$. This indicates stronger compression than inferred from the density at the peak velocity. In contrast, Parcel~2, launched from the center of the trailing boundary region, decreases in density from $0.57$ to $0.39$, i.e., by \SI{32}{\percent}, corresponding to $\alpha = -2.27$. This depletion results from the dispersing velocity distribution in the stream tail, which leads to an enhanced radial expansion. This contrasts with the density evolution evaluated at the minimum density of the stream, which indicated compression in the high-speed stream tail rather than the depletion experienced by Parcel~2.

The discrepancy between the density evolution of individual plasma parcels and that inferred from prominent locations within the high-speed stream arises because these locations are not tied to fixed plasma elements but instead evolve dynamically within the stream. The peak velocity corresponds to the fastest plasma parcels at a given time; however, as the stream interaction region progressively decelerates the leading plasma, the set of parcels defining the peak velocity changes continuously. Similarly, the minimum density does not remain associated with the original stream core. As tail plasma disperses and expands, its density decreases and eventually falls below that of the core, shifting the location of the minimum density toward the tail. Therefore, in general, the density evolution evaluated at prominent locations in the high-speed stream does not reflect the density evolution of the plasma parcels. Only in the extended plateau of a very large high-speed stream would the density evolution at a fixed location reliably trace that of the plasma parcels. This is restricted to those parts of the plateau that are neither affected by any pressure gradients associated with the stream interaction region nor by the velocity dispersion toward the stream tail. Such large high-speed streams, however, are rarely encountered in observations.

The temperature evolution of the high-speed stream is shown in Figure~\ref{fig:radial_vnt}(c). To isolate deviations from adiabatic cooling, we first remove the expected decrease in temperature due to the spherical expansion of the plasma, which follows a geometric $r^{4/3}$ scaling, and then normalize the resulting temperature to the ambient slow solar wind temperature at the inner boundary. Near the Sun, the normalized temperature distribution remains roughly symmetric across the leading and trailing edges of the high-speed stream, mirroring the density and velocity distributions. As the stream propagates outward, the temperature steepens at the leading edge, reaches a pronounced peak, and then decreases more gradually toward the stream tail. This steep increase at the front arises from the thermalization of high-speed stream plasma impinging on the stream interaction region.

This behavior is also reflected in the evolution of the tracked plasma parcels. Parcel~1, originating from the stream core, experiences a sharp temperature increase as it is decelerated by the pressure gradient in the stream interaction region. In contrast, Parcel~2, originating from the center of the trailing boundary region, shows only a modest temperature rise while propagating through interplanetary space. This increase arises from a small, artificial heating in the EUHFORIA code, applied to reproduce the residual solar wind acceleration up to \SI{0.4}{AU}. This artificial increase might also mask an expected, slight temperature decrease in the high-speed stream tail due to radial expansion. We do not attempt to fit a power-law dependence for the temperature evolution, as any such fit would be influenced by this artificial heating.

These dynamics have important implications for interpreting solar wind observations. 
Observational studies typically analyze the evolution of high-speed streams by comparing measurements from two spacecraft at different heliocentric distances. Since individual plasma parcels cannot be tracked in practice, analyses often focus on easily identifiable features, such as the evolution of the peak velocity, the density at the peak velocity or in the velocity plateau, or the minimum density in the high-speed stream. However, these features do not correspond to the same plasma parcels as the stream propagates outward. Consequently, analyses based on such fixed features do not follow the same plasma, therefore do not accurately represent the underlying plasma evolution, and consequentially do not allow for meaningful power-law fits describing that plasma evolution.

Fitting the plasma evolution to plasma properties averaged over large high-speed stream areas or even large ensembles of high-speed streams is similarly dangerous, as the evolution of plasma parcels strongly depends on their starting conditions on the Sun. It makes a large difference for the density evolution whether a plasma parcel with a constant speed of \SI{600}{km\,s^{-1}} originates from the high-speed stream core or from the trailing boundary where a velocity gradient is present, causing an additional radial expansion of the plasma parcels while they propagate through interplanetary space. Averaging over such datasets without accounting for these differing origins can therefore introduce systematic biases, depending on how the dataset is composed in terms of contributions from stream cores and boundary regions.

\subsection{Magnetic field}

\begin{figure}[t!]
 \includegraphics[width=\textwidth]{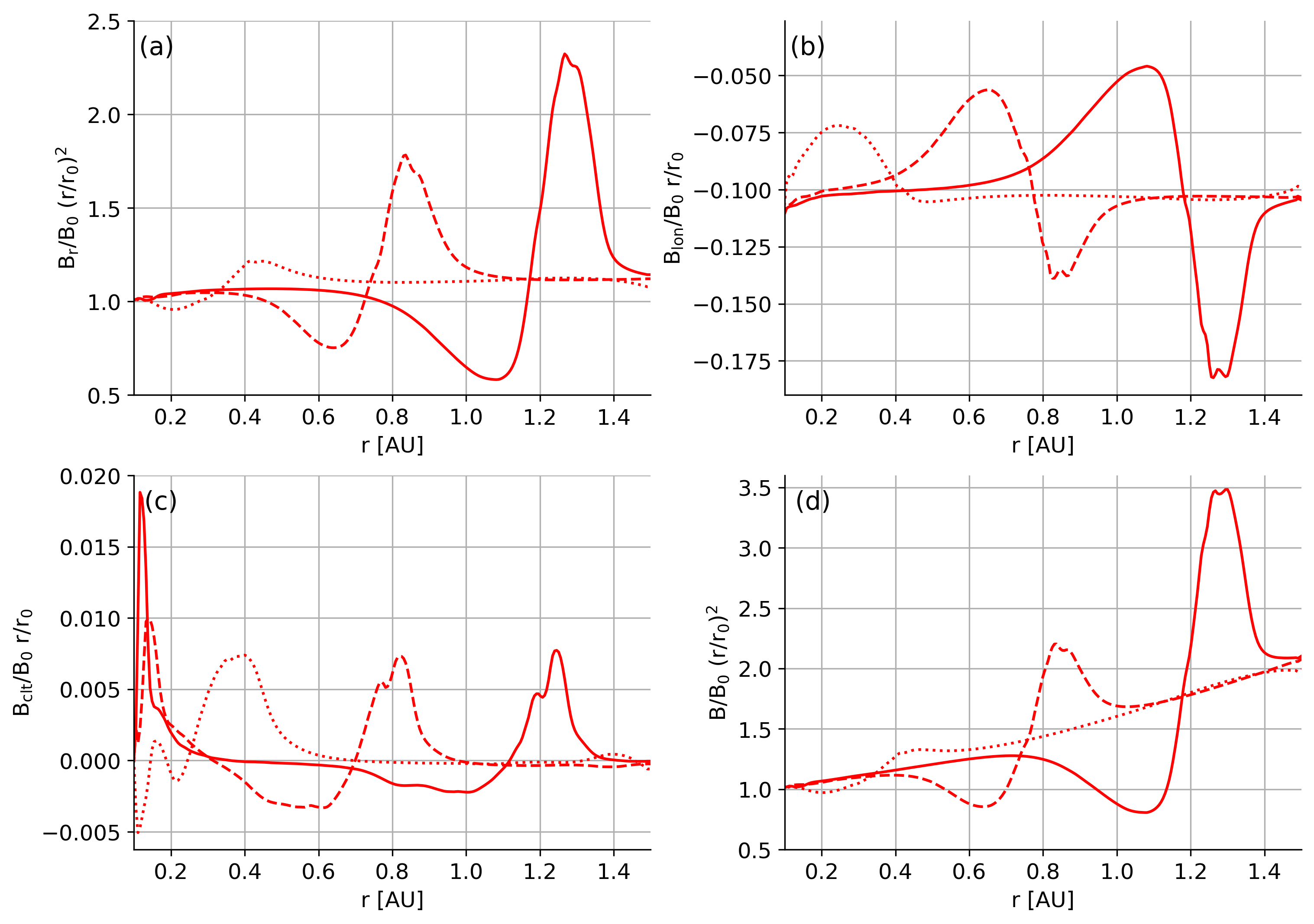}
\caption{Evolution of high-speed stream properties at the stream center. Panels show (a)~the radial, (b)~longitudinal, (c)~colatitudinal, and (d)~total magnetic field strengths along a fixed radial direction after $1$~(dotted), $2.5$~(dashed), and $4$~days. The radial and total components have been scaled by $(r/r_0)^2$, while the longitudinal and colatitudinal components have been scaled by $r$ to remove the expected effects of spherical expansion and magnetic field rotation with increasing distance from the Sun. All scaled quantities are further normalized by the ambient slow solar wind magnetic field strength $B_0$ at the inner boundary, expressing the field strengths relative to near-Sun slow solar wind conditions while preserving the relative amplitudes of the individual components to each other.}
\label{fig:radial_B}
\end{figure}

We investigate the evolution of the radial, longitudinal, and co-latitudinal components of the magnetic field, as well as the total magnetic field strength, in high-speed streams in Figure~\ref{fig:radial_B}. Unless stated otherwise, we evaluate the magnetic field at the latitudinal center of the high-speed stream. The exception is the co-latitudinal component, which we analyze at a latitude of \SI{6}{\degree} relative to the stream center. This is because the co-latitudinal component arises from latitudinal plasma deflection at the stream interface and therefore vanishes at the stream center, as will be discussed in Section~\ref{sec:si}.

For each component, we remove the corresponding geometric scaling with radial distance, as discussed below, and normalize the magnetic field to the absolute magnetic field strength of the ambient solar wind at the inner boundary. This normalization allows for a direct comparison of how the magnetic field is distributed among its spatial components. In general, we expect the magnetic field evolution to follow the density evolution, as the magnetic field is frozen into the solar wind plasma.

The evolution of the radial magnetic field component is shown in Figure~\ref{fig:radial_B}(a). We remove the geometric $r^2$ decrease associated with spherical expansion. In this normalized representation, each snapshot shows a strong enhancement of the magnetic field in the stream interaction region due to compression, and a depletion in the stream tail caused by radial expansion. Both effects become more pronounced with increasing distance from the Sun. 

To assess the conservation of radial magnetic flux, we integrate the radial magnetic field as the high-speed stream passes three virtual spacecraft located at $0.2$, $0.6$, and~\SI{1}{AU}. Because the virtual spacecraft samples only a longitudinal slice through the stream and its latitudinal extent is not known, this integration mimics in-situ observations and provides a measure of the magnetic flux based solely on such a slice. We find that the radial magnetic flux derived in this way is well conserved with distance. Thus, compression and depletion redistribute the radial magnetic field within the stream, but do not change its total flux.

Next, we consider the longitudinal magnetic field component, shown in Figure~\ref{fig:radial_B}(b). Here, we remove a linear $r$-dependence, reflecting the fact that the longitudinal component arises from the rotation of the originally radially-orientated magnetic field into the Parker spiral, so that the longitudinal magnetic field component increases approximately linearly with distance from the Sun. A negative longitudinal component represents a westward magnetic field direction, while a positive one would represent an eastward direction. The longitudinal field is consistently negative, confirming a westward orientation along the Parker spiral. As for the radial component, compression in the stream interaction region enhances the magnitude of the longitudinal component (toward more negative values), while expansion in the tail reduces its average magnitude (towards less negative values).

Although a longitudinal deflection of the magnetic field at the stream interface would be expected—since the field is frozen into the plasma and the longitudinal velocity shows such a deflection (see Figure~\ref{fig:hss_definition}(d))—this effect is not clearly visible. This likely reflects that the deflection is small compared to the dominant Parker spiral rotation, but may also indicate a limitation of MHD simulations, which tend to underestimate deflection effects in stream interaction regions.

The co-latitudinal magnetic field component is shown in Figure~\ref{fig:radial_B}(c). We remove a linear $r$-dependence, although the physical origin of this scaling is not fully clear. This component is evaluated at \SI{6}{\degree} from the stream center, where latitudinal deflection at the stream interface is significant. The co-latitudinal field is about an order of magnitude weaker than the longitudinal component. However, its magnitude may be underestimated due to the limited deflection captured in the simulations.

Similar to the other components, the co-latitudinal magnetic field shows compression in the stream interaction region and depletion in the tail. Notably, the compression region exhibits a double-peaked structure. The outer peak, at larger radial distances, arises from slow solar wind plasma being deflected away from the stream center, while the inner peak, at smaller radial distances, results from high-speed stream plasma being deflected back toward the center at the rear of the interface. In our simulations, the high-speed stream and the preceding slow solar wind have the same magnetic polarity, so the co-latitudinal component of these two peaks in the stream interaction region maintains the same sign. If the high-speed stream and preceding slow solar wind had opposite magnetic polarities, the peaks would exhibit opposite signs, producing a zero-crossing in the latitudinal field between the two peaks.

The total magnetic field strength is shown in Figure~\ref{fig:radial_B}(d). After removing the geometric $r^2$ decrease expected from spherical expansion, we again observe clear compression and depletion regions. However, the total field strength exhibits an overall increase with distance in this normalized representation. This arises because the radial component scales as $r^{-2}$, while the longitudinal component grows linearly with distance and increasingly dominates the total field.
When integrating the total magnetic field strength along the spacecraft trajectories at $0.2$, $0.6$, and~\SI{1}{AU}, we find that this measure for the associated magnetic flux increases by \SI{44}{\percent} between $0.2$ and \SI{1}{AU}. Thus, unlike the radial component, the integral of the total magnetic field strength is not conserved as the stream propagates outward.

This difference can be understood from the definition of magnetic flux. Magnetic flux is given by the integral of the magnetic field component perpendicular to a surface. A spacecraft at a fixed heliocentric distance samples the solar wind stream along a path that corresponds to a longitudinal cut on a spherical surface. The normal direction of this surface is radial, so the radial magnetic field component is perpendicular to the sampling surface. Consequently, integrating the radial component along a longitudinal cut yields a conserved flux.

In contrast, the total magnetic field is oriented along the Parker spiral and is therefore not perpendicular to the sampling surface. To derive the corresponding magnetic flux, one would need to integrate over a surface perpendicular to the Parker spiral direction. This surface is inclined relative to the spherical surface, and the inclination angle increases with distance from the Sun. Consequently, to correctly compute a conserved magnetic flux from the total magnetic field, a correction factor is needed to account for this changing geometry with increasing distance from the Sun.

From this follows that when estimating the solar magnetic flux that is open to interplanetary space from in-situ measurements, the radial magnetic field component should be used rather than the total field. Near the Sun, the solar magnetic field is oriented radially, and the radial component is conserved as the high-speed stream propagates outward. Therefore, the integral of the radial magnetic field component measured by a spacecraft at a fixed distance from the Sun provides a correct estimate of the solar magnetic flux that is open to interplanetary space.

\subsection{Particle Flux and and kinetic energy flux}

\begin{figure}[t!]
 \includegraphics[width=.5\textwidth]{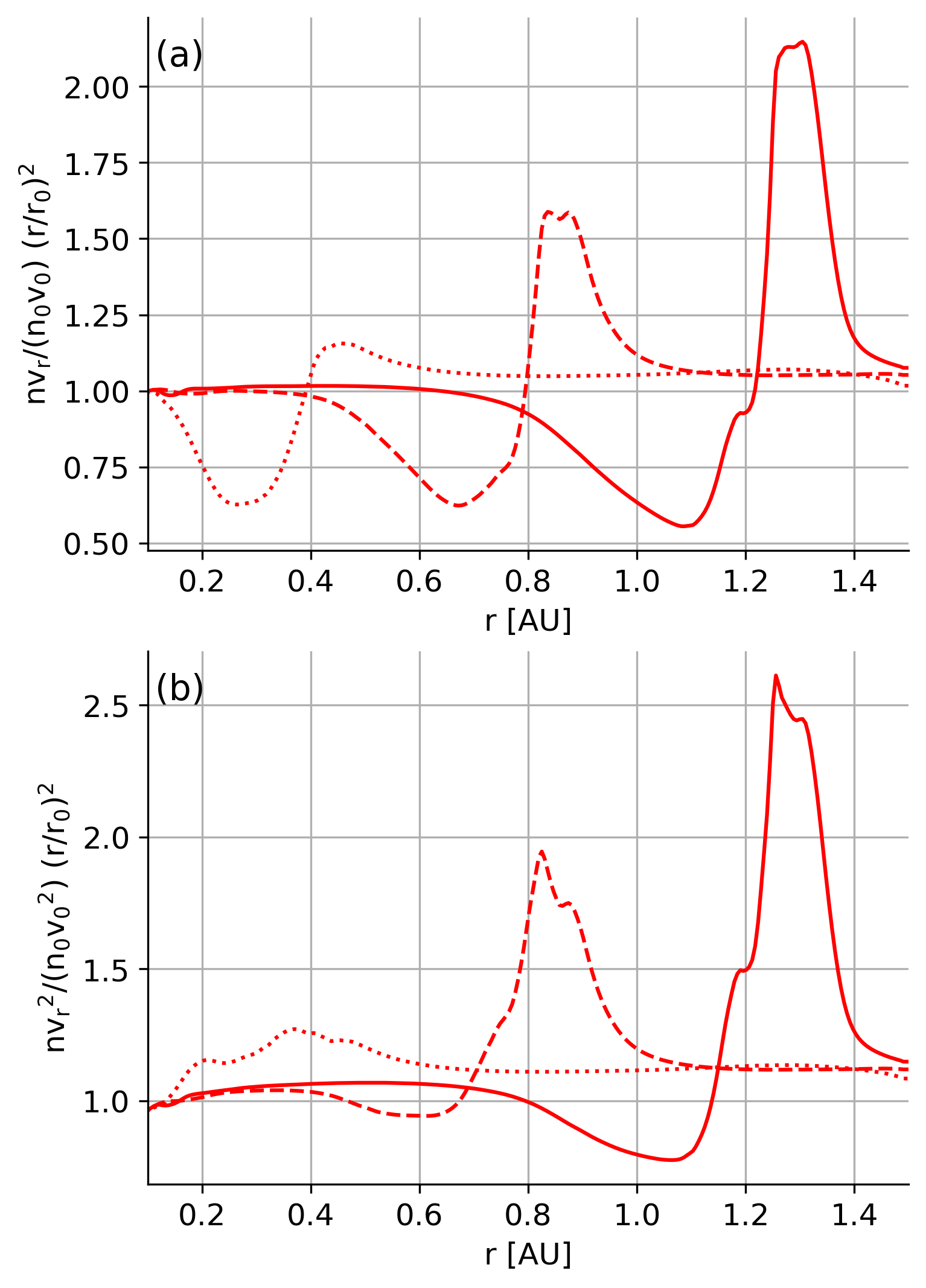}
\caption{Evolution of high-speed stream properties at the stream center. Panels show (a)~the particle flux and (b)~the kinetic energy density along a fixed radial direction after $1$~(dotted), $2.5$~(dashed), and $4$~days. The particle flux and kinetic energy densities have been scaled by $(r/r_0)^2$ to remove the expected effects of spherical expansion. The scaled quantities are further normalized by the ambient slow solar wind particle flux and kinetic energy density, respectively, expressing the stream properties relative to near-Sun slow solar wind conditions.}
\label{fig:nv}
\end{figure}

We investigate the radial evolution of the particle flux, $n\,v$, and the kinetic energy density, $0.5\,n\,v^2$, which are shown in Figure~\ref{fig:nv}. We again remove a geometric $r^2$ scaling related to the radial expansion of the plasma and normalize the quantities to their respective values of the ambient solar wind at the inner boundary. 

The evolution of the particle flux is shown in Figure~\ref{fig:nv}(a). Since the velocity of a plasma parcel is about constant but its density changes due to compression in the stream interaction region and expansion in the tail, the particle flux resembles the evolution of these quantities: with increasing distance from the Sun, the particle flux increases with distance from the Sun in the stream interaction region due to compression and pile-up of plasma, while the particle flux in the high-speed stream tail decreases due to radial expansion. When averaging the particle flux as seen by virtual satellites at $0.2$,~$0.6$, and~\SI{1}{AU} over the high-speed stream, i.e., from the tail to the stream interface, we find that the average mass flux slightly increases from $0.2$~to~\SI{1}{AU} by about \SI{5}{\percent}. This slight increase might be related to residual solar wind acceleration under non-steady conditions in interplanetary space, such as those associated with the passing high-speed stream.

The evolution of the kinetic energy density is shown in Figure~\ref{fig:nv}(b). It follows the same general behaviour as the particle flux, with enhanced peaks in the stream interaction region and a progressively stronger depletion in the high-speed stream tail at larger heliocentric distances. When averaging over the high-speed stream interval, we find a decrease in the kinetic energy density of about \SI{7}{\percent} from $0.2$ to \SI{1}{AU}. This decrease reflects the deceleration of the fastest plasma parcels in the stream interaction region. In contrast, integrating over the full high-speed stream interval including the stream interaction region yields an increase of about \SI{92}{\percent} in the total kinetic energy. This increase in kinetic energy reflects the accumulation of plasma in the compressed stream structure due to the continuous pile-up of the slow solar wind. 

This increase in the total kinetic energy associated with slow solar wind pile-up has implications for the geoeffectivity of high-speed streams. When the stream interaction region reaches Earth’s magnetosphere, the solar wind dynamic pressure, which is proportional to the kinetic energy density, compresses the magnetosphere and reduces its standoff distance, thereby modifying its global shape and geometry. The enhanced kinetic energy content resulting from the accumulated slow solar wind thus substantially contributes to the overall compressive impact of the stream interaction region.

This increase in kinetic energy  with radial distance from the Sun by slow solar wind pile-up has an effect on the geoeffectivity of high-speed streams. When the stream interaction region hits Earth's magnetosphere, the high-speed stream's dynamic pressure, which is equivalent to the kinetic energy density, compresses Earth's magnetosphere and thereby decreases its distance from Earth's surface, changing the geometric extension of the magnetosphere. The slow solar wind pile-up substantially increases the kinetic energy causing the compression and thus has an important effect here.

\subsection{Relation between velocity, density, and magnetic field}

\begin{figure}[t!]
 \includegraphics[width=\textwidth]{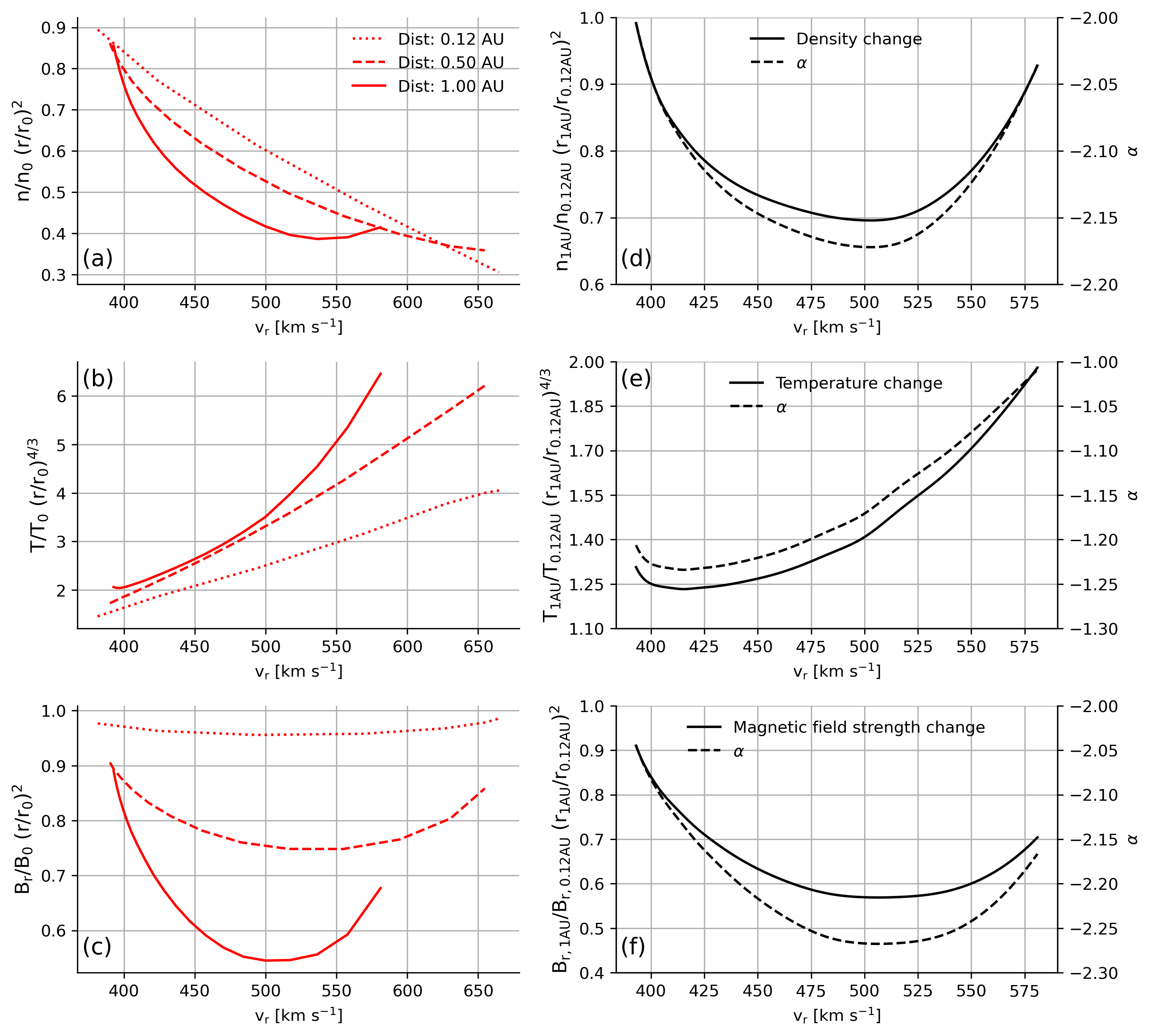}
\caption{Relationships between high-speed stream parameters evaluated at distances of $0.12$, $0.5$, and \SI{1}{AU} from the Sun. Panels~(a)–(c) show density vs. velocity, temperature vs. velocity, and radial magnetic field strength vs. velocity, respectively. Panels~(e)–(g) show the change in density, temperature, and radial magnetic field strength between $0.12$ and \SI{1}{AU} as a function of velocity, together with polynomial fits for their radial decrease of the form $x_r = x_0\,r^\alpha$, where $x$ denotes density, temperature, and magnetic field strength, respectively. 
All densities and temperatures have been scaled by $(r/r_0)^2$ and $(r/r_0)^{4/3}$, respectively, to remove the expected effects of spherical expansion. For panels~(a)–(c), the quantities are further normalized by the ambient slow solar wind density $n_0$ and temperature $T_0$ at the inner boundary, expressing them relative to near-Sun slow solar wind conditions. The polynomial exponents were derived from the unscaled variables.
}
\label{fig:vnTB}
\end{figure}

We investigate the evolution of the velocity--density, velocity--temperature, and velocity--magnetic field relationships with distance from the Sun. To isolate physical effects, we remove the geometric $r^2$ scaling for the density and magnetic field, and the geometric $r^{4/3}$ scaling for the temperature that arises from adiabatic cooling due to spherical expansion. We then normalize all quantities to their respective ambient slow solar wind values at the inner boundary. We use three virtual spacecraft located at $0.12$, $0.5$, and~\SI{1}{AU} to sample the plasma properties as the high-speed stream passes. In this analysis, we exclude the stream interaction region. The resulting relationships are shown in Figure~\ref{fig:vnTB}.

We first consider the velocity--density relationship in Figure~\ref{fig:vnTB}(a). Close to the Sun, this relationship is nearly linear, reflecting the transition across the boundary layer from the ambient slow solar wind to the high-speed stream, where the velocity increases while the density decreases at a similar rate. As the stream propagates outward, the density in the high-speed stream tail decreases more rapidly than in the core due to radial expansion driven by the velocity gradient in the tail. This gradient is strongest in the center of the trailing boundary region and weakest near the ambient solar wind and the stream core. Consequently, the initially linear relationship develops a pronounced bulge with increasing distance from the Sun, with the bulge's center corresponding to plasma parcels from the trailing boundary region that experience the strongest radial expansion. At \SI{1}{AU}, the highest-velocity parcels have been partially absorbed by the stream interaction region, which reduces the peak velocity in the distribution.

The velocity--temperature relationship, shown in Figure~\ref{fig:vnTB}(b), exhibits a similar evolution. Close to the Sun, the relationship is again approximately linear, reflecting the transition across the boundary layer from the ambient slow solar wind to the high-speed stream plasma. With increasing distance from the Sun, the slope of the relationship increases and a bulge develops. The steeper slope results from the residual heating implemented in the EUHFORIA code, which drives the continued acceleration of the solar wind in interplanetary space. The bulge, in contrast, again reflects the enhanced radial expansion in the high-speed stream tail, which leads to stronger adiabatic cooling in this region.

The velocity--magnetic field strength relationship is shown in Figure~\ref{fig:vnTB}(c). In contrast to the other relationships, it is nearly flat close to the Sun. This behavior results directly from our initialization, in which we impose magnetic pressure equilibrium between the high-speed stream and the surrounding ambient slow solar wind. This setup reflects the state reached after the radial expansion of the high-speed stream in the low-$\beta$ corona, where such an equilibrium is expected. With increasing distance from the Sun, the relationship develops a bulge, which arises from radial expansion in the high-speed stream tail and the associated radial expansion of the frozen-in magnetic field, leading to a reduction in the magnetic field strength in this region.

The evolution of these relationships with heliospheric distance offers diagnostic potential for studying the radial evolution of high-speed streams. As discussed in Section~\ref{sec:vnt}, analyzing the radial evolution of density and temperature directly is challenging, because the global structure of the stream evolves with distance and does not provide a stable reference frame for tracking plasma parcels. However, we have shown that, outside the stream interaction region, the velocity of plasma parcels remains nearly constant. The velocity can therefore serve as a proxy to relate plasma parcels observed at different heliocentric distances.

Using this approach, we find from Figure~\ref{fig:vnTB}(a) that plasma parcels with velocities around \SI{500}{km\,s^{-1}}, originating from the center of the trailing boundary region, decrease in density from $0.60$ at \SI{0.15}{AU} to $0.41$ at \SI{1}{AU}. This corresponds to an additional radial expansion of \SI{32}{\percent} beyond the geometric $r^2$ decrease, in good agreement with the \SI{34}{\percent} decrease derived from tracking plasma parcels in Section~\ref{sec:vnt}. In contrast, plasma parcels at \SI{390}{km\,s^{-1}}, originating from the trailing high-speed stream boundary near the ambient slow solar wind, exhibit nearly unchanged normalized densities between \SI{0.12}{AU} and \SI{1}{AU} in Figure~\ref{fig:vnTB}(a), indicating only minimal radial expansion beyond spherical expansion. The velocity--temperature and velocity--magnetic field relationships can be analyzed in an analogous manner, providing a new method to study the radial evolution of high-speed stream plasma parcels.

[Show the density reduction, temperature increase, and magnetic field reduction as an additional line in the plots and refer here to it.]

\section{The stream interface between HSSs and the preceding slow solar wind} \label{sec:si}

\begin{figure}[t!]
 \includegraphics[width=\textwidth]{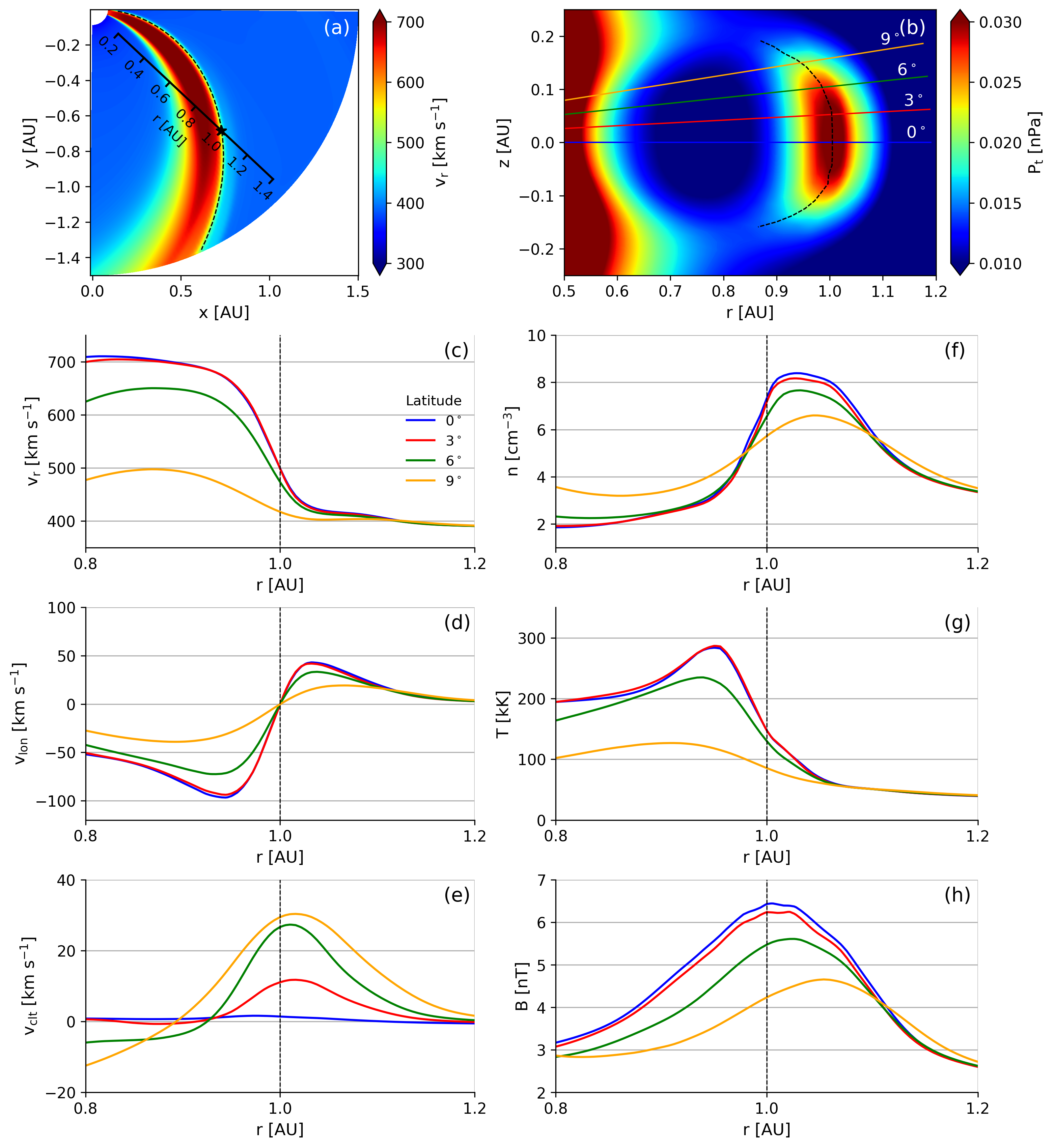}
\caption{Stream properties across the stream interaction region. Panel~(a) shows the stream velocity in the heliospheric equatorial plane, and panel~(b) the transverse pressure in the radial–latitudinal plane. The stream interface is indicated by the dashed black line in each panel; in panel~(b), four latitudinal directions at \SI{0}{\degree}, \SI{3}{\degree}, \SI{6}{\degree}, and \SI{9}{\degree} are marked. Panels~(c)–(h) are evaluated at the time when the stream interface, marked by the dashed black lines, crosses \SI{1}{AU}, and show the radial, longitudinal, and colatitudinal velocity components, as well as the density, temperature, and total magnetic field strength, as functions of radial distance for each of the four latitudinal directions across the stream interface.}
\label{fig:sir_overview}
\end{figure}

\begin{figure}[t!]
 \includegraphics[width=\textwidth]{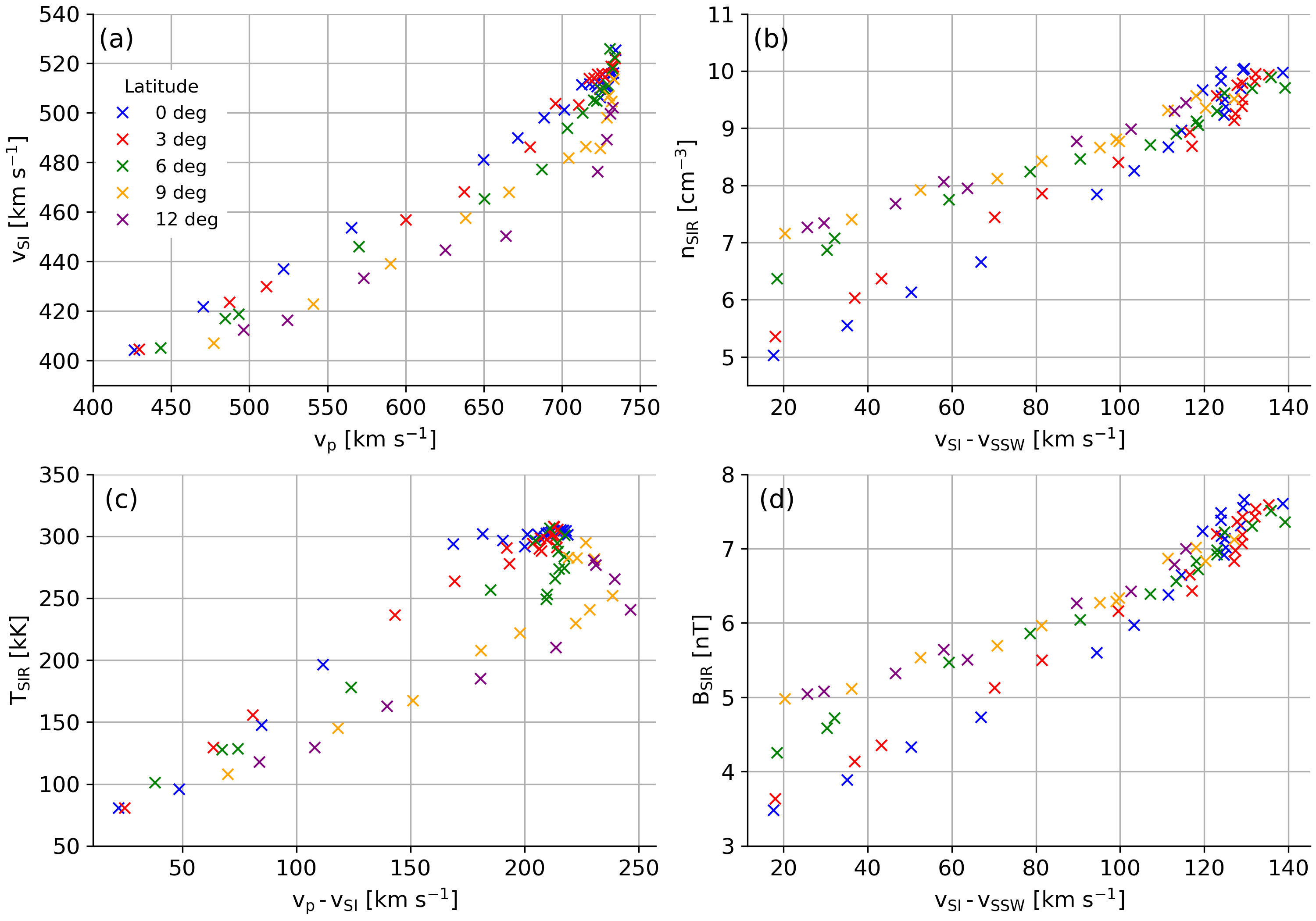}
\caption{Relationships between stream interaction region properties. (a)~Stream interface velocity vs. stream peak velocity. (b)~Peak density in the stream interaction region vs. the velocity difference between the stream interface and the preceding ambient slow solar wind, governing slow solar wind pile-up and compression. (c)~Peak temperature in the stream interaction region vs. the velocity difference between the stream peak velocity and the stream interface velocity, reflecting the energy available for thermalization. (d)~Peak magnetic field strength in the stream interaction region vs. the velocity difference between the stream interface and the preceding slow solar wind, governing magnetic field pile-up and compression.}
\label{fig:sir_relations}
\end{figure}

The stream interface is a two dimensional surface, and the associated stream interaction region a three dimensional volume that separates the high speed stream from the preceding slow solar wind. They extend along the Parker spiral in the equatorial plane and along the high speed stream front in the latitudinal direction, as shown in Figure~\ref{fig:sir_overview}(a) and~(b). In the following, we first examine the properties of the stream interaction region across its latitudinal front and then analyze the relationships between its properties.

\subsection{Latitudinal plasma distributions}

The transverse pressure, defined as the sum of gas and magnetic pressure, characterizes the large scale structure of the stream interaction region and is shown in Figure~\ref{fig:sir_overview}b. The stream interface and stream interaction region appear as a region of enhanced pressure to the right of the image center, followed by the low pressure cavity of the high speed stream to the left. These pressure gradients drive the plasma dynamics by accelerating, decelerating, and deflecting plasma parcels.

At the front of the stream interface, the pressure decreases toward larger radial distances and increasing height above the equatorial plane. This gradient accelerates the slow solar wind radially while deflecting it away from the equatorial plane, so that it streams slowly along the latitudinal extent of the stream interface away from its center. Behind the stream interface, within the stream interaction region, the pressure gradient reverses and increases toward the stream interface center. This decelerates impinging high-speed stream plasma and drives a latitudinal deflection away from the center, resulting in a slow outflow from the central stream interaction region.
Only deeper inside the high speed stream, the pressure gradient changes orientation and increases toward higher latitudes. This results in a small deflection of high speed stream plasma toward the latitudinal center of the stream, leading to a net convergence of the high-speed stream behind the stream interaction region toward the stream center.

In the following panels~(c)--(h), we show the plasma distributions along four latitudinal directions at $0$, $3$, $6$, and \SI{9}{\degree}, sampling distributions from the stream center to its northern flank. The radial velocity distribution of the solar wind across the stream interaction region is shown in panel~(c). The peak velocity decreases from \SI{700}{km,s^{-1}} in the stream center to \SI{500}{km,s^{-1}} in the flanks. This variation mainly reflects that plasma in the flanks originates from the latitudinal boundary region of the high speed stream close to the Sun, where velocities are lower.

The velocity of the stream interface is a result from momentum conservation between the impinging high speed stream and the preceding piled up slow solar wind. From panel~(c), it is apparent that it remains well below the peak velocity of the high speed stream, varying from \SI{500}{km,s^{-1}} in the stream center to \SI{430}{km,s^{-1}} in the flanks. This corresponds to a velocity difference between the peak velocity of the high-speed stream and the velocity of the stream interface of \SI{200}{km\,s^{-1}} in the stream center and only \SI{70}{km\,s^{-1}} in the stream flank. The velocity difference between the stream interface and the preceding slow solar wind of \SI{100}{km,s^{-1}} in the stream center and only \SI{30}{km,s^{-1}} in the flank.

The longitudinal velocity component, which indicates east-west deflection, is shown in panel~(d). In front of the stream interface, the component is positive, indicating a westward deflection of the slow solar wind away from the Parker spiral direction. Behind the stream interface, the component becomes negative, indicating an eastward deflection of the impinging high-speed stream along the Parker spiral. The longitudinal deflection is strongest in the stream center and decreases toward the flanks, consistent with the impingement speed of plasma into the stream interaction region.

The colatitudinal velocity component, which indicates north-south deflection, is shown in panel~(e). In front of and behind the stream interface, the component is positive, indicating flow toward solar north and thus away from the stream center. Only deeper inside the high speed stream does the flow acquire a weak southward component, indicating a slight convergence toward the stream center. The colatitudinal deflection is strongest in the flank, where the stream interface is most inclined relative to the radial direction, and negligible in the center, so that the stream remains nearly radial there.

The plasma density across the stream interaction region is shown in panel~(f). Here, we did not remove a geometrical factor due to the spherical expansion. The compression due to piled-up slow solar wind plasma and accumulating high speed stream plasma is clearly visible. In a one dimensional picture, neglecting latitudinal effects, the pile up depends mainly on the velocity difference between the stream interface and the preceding slow solar wind \citep{hofmeister2022}. With a three-times higher velocity difference in the stream center as compared to the flanks, 0ne would expect significantly stronger pile up in the stream center than in the flanks. However, the simulations show only about \SI{25}{\percent} higher density in the center than in the flanks. The reason for this rather small difference is latitudinal plasma transport in the stream interaction region away from its center, which redistributes mass towards the flanks. 

The plasma temperature is shown in panel~(g).  After the stream interface, the temperature rises steeply due to thermalization of high speed stream plasma as it impinges on the stream interaction region. The peak temperature in the stream center is about a factor of 2.5 higher than in the flanks. This results from the velocity difference between the peak velocity of the high speed stream and the stream interface, which is much larger in the stream center than in the flanks, and which determines the available energy for thermalization and thus heating.

The magnetic field strength is shown in panel~(h). Similar to the density, a clear compression is visible, with about \SI{25}{\percent} stronger magnetic fields in the stream center than in the flanks. Based on the velocity difference between the stream interface and the preceding slow solar wind, one would expect a stronger compression in the stream center and therefore a larger contrast in magnetic field strength. However, because the magnetic field is frozen into the solar wind plasma, the latitudinal flows along the stream interface away from the stream center towards the flanks also redistribute the magnetic flux, which reduces the expected difference.

These results lead to two main conclusions. First, latitudinal flows at the stream interface redistribute both mass and magnetic field from the stream center toward the flanks, which requires fully three dimensional MHD modeling for quantitative studies. Second, the colatitudinal velocity component provides a geometric diagnostic: it points away from the stream center in the interaction region and toward it within the high speed stream. In situ measurements can therefore determine whether the stream center passes north or south of the spacecraft and, assuming near radial propagation, constrain whether the source coronal hole lies in the northern or southern solar hemisphere.

\subsection{Relationships between parameters}

We investigate the relationships between the peak velocity of the high speed stream, the velocity of the stream interface, and the peak density, temperature, and magnetic field strength in the stream interaction region in Figure~\ref{fig:sir_relations}. In this figure, the dataset comprises all high-speed stream diameters with measurements at all latitudies in our simulations.

The velocity of the stream interface is set by momentum conservation between the impinging high speed stream plasma and the preceding slow solar wind plasma \citep{hofmeister2022}. It therefore depends on the peak velocity and density of the high speed stream and on the properties of the preceding slow solar wind. In the simulations, the slow solar wind is kept constant, and the high speed stream density is coupled to the stream velocity through the latitudinal origin of the plasma near the Sun. As a result, the stream interface velocity depends primarily on the peak high speed stream velocity, as shown in Figure~\ref{fig:sir_relations}(a). This dependence is approximately linear, with slightly higher velocities of the stream interface in the stream center than at higher latitudes. This latitudinal difference results from the latitudinal redistribution of piled up slow solar wind mass across the stream interaction region. This plasma transport from the center to the flanks results in a modified momentum balance, which reduces the velocity of the stream interface in the flanks.

The peak density in the stream interaction region is a function of the velocity difference between the stream interface and the preceding slow solar wind, as shown in Figure~\ref{fig:sir_relations}(b). The density increases approximately linearly with this velocity difference. At a constant small velocity difference, such as \SI{30}{km,s^{-1}}, the peak density is about \SI{30}{\percent} higher in the high speed stream flanks than in the center. This difference arises from the latitudinal mass transport within the stream interaction region. At larger velocity differences between the stream interface and the preceding slow solar wind, the stronger pile up dominates and reduces the relative impact of latitudinal transport.

The peak temperature depends on the velocity difference between the peak velocity of the high speed stream and the velocity of the stream interface, as shown in Figure~\ref{fig:sir_relations}(c). This difference determines the available kinetic energy for thermalization and thus heating. The temperature increases approximately linearly with this velocity difference and shows more scatter at higher velocity differences. 

The peak magnetic field strength in the stream interaction region, shown in Figure~\ref{fig:sir_relations}(d), depends again on the velocity difference between the stream interface and the preceding slow solar wind. This scaling mirrors the density behaviour with higher magnetic field strength at higher velocity differences, because the magnetic field is frozen into the plasma.

It is noteworthy that the velocity of the stream interface depends on the peak velocity of the high speed stream. Therefore,  all relevant velocity differences, and as a result also peak density, peak magnetic field strength, and peak temperature are linear functions of the peak velocity of the high-speed stream.

\section{Properties and geoeffectivity \SI{1}{AU}} \label{sec:1au}

In this section, we combine the information learned in the previous sections to investigate the properties and geoeffectivity of high-speed streams at \SI{1}{AU}. The following stream properties are derived from measurements of a virtual spacecraft at a distance of \SI{1}{AU} from the Sun and at varying latitudinal angles between the high-speed stream centers and their flanks. To assess geoeffectivity, we assume Earth to be located at a given latitudinal offset from the stream center and account for the seasonal variation in the orientation of Earth's magnetic field axis.

\subsection{Properties at \SI{1}{AU}} \label{sec:1au}

\begin{figure}[t!]
 \includegraphics[width=\textwidth]{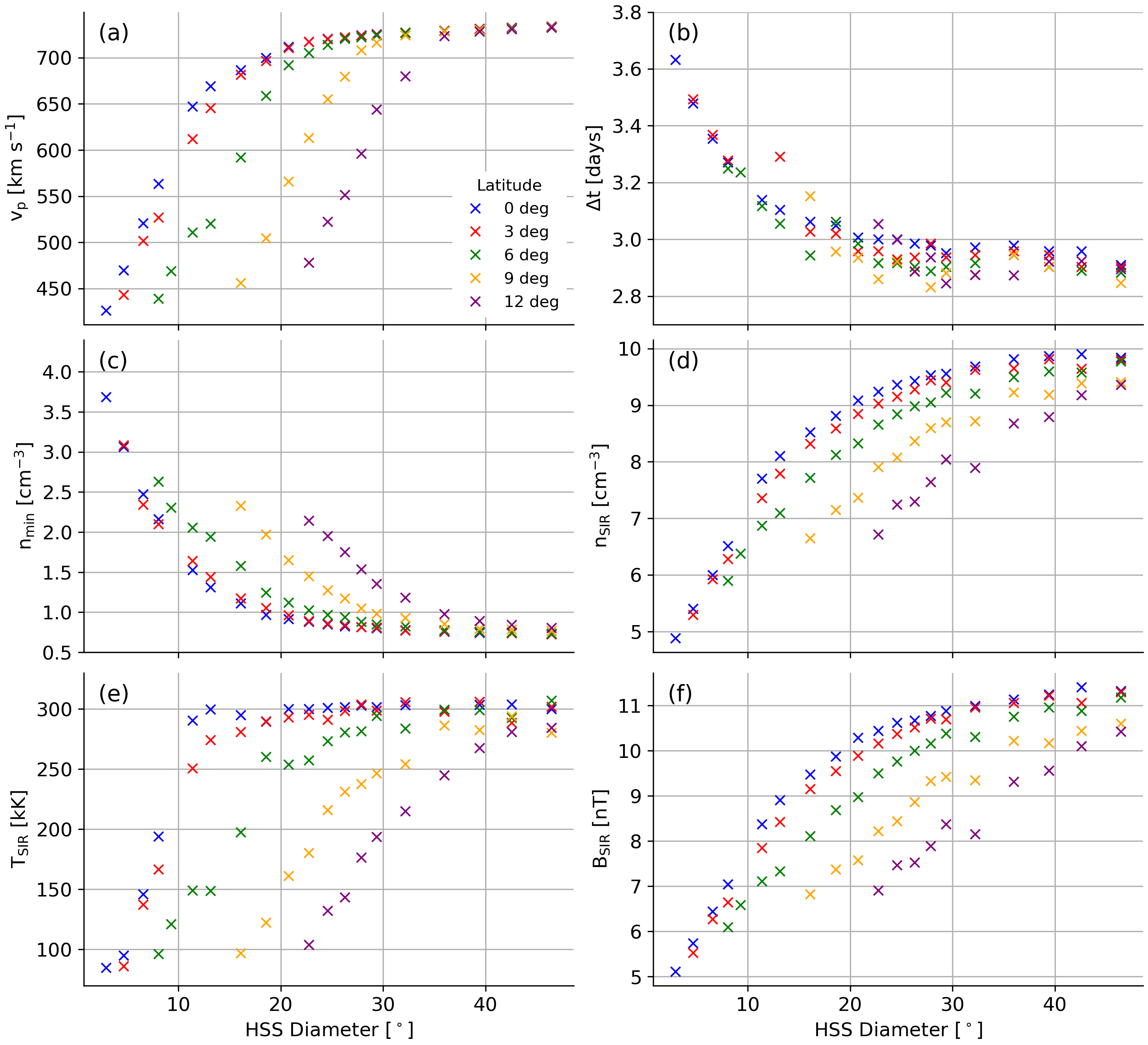}
\caption{High-speed stream characteristics at \SI{1}{AU}. (a)~Peak velocity, (b)~travel time, (c)~minimum density, (d)~peak density, (e)~peak temperature, and (f)~peak magnetic field strength as functions of the stream diameter near the Sun and the latitudinal position of the measurement within the stream.}
\label{fig:hss_1au}
\end{figure}

We investigate the global properties of high-speed streams at \SI{1}{AU}, and analyze how locally measured properties deviate from these global values, depending on the stream diameter near the Sun and the spacecraft’s latitudinal position within the stream. The results are presented in Figure~\ref{fig:hss_1au}.

The global peak velocity of a stream with a given diameter at \SI{1}{AU} occurs at its center, as shown in panel~(a). It increases approximately linearly with stream diameter up to about \SI{14}{\degree}, after which the increase slows and eventually saturates at \SI{730}{km,s^{-1}}. In contrast, the local peak velocity, measured by a virtual spacecraft within a fixed-diameter stream, decreases significantly with increasing latitude away from the stream center toward the flanks. All streams have been initialized with identical velocities at the inner boundary.

The reduction of the global peak velocity for small stream diameters is a consequence of boundary-region effects and interplanetary deceleration. In all simulations, the core velocity at the inner boundary is set to \SI{650}{km\,s^{-1}}, corresponding to a maximum possible velocity of about \SI{730}{km\,s^{-1}} at \SI{1}{AU} when including residual acceleration. For small streams with diameters \SI{<14}{\degree}, however, the entire stream is dominated by boundary-region plasma, which already exhibits reduced velocities close to the Sun (Section~\ref{sec:closeSun}). In addition, as the stream propagates outward, the fastest plasma parcels from the core and later from the trailing boundary progressively impinge on the stream interaction region and are decelerated (Section~\ref{sec:vnt}), reducing the peak velocity. Together, these effects reduce the global peak velocity despite identical initial plasma conditions in the simulations.

A similar mechanism explains the latitudinal dependence of the local peak velocity measured by a satellite. With increasing latitude, the spacecraft samples regions closer to the stream’s boundary layers rather than its core. As a result, the measured peak velocity becomes increasingly influenced by boundary-region plasma, which already exhibits reduced velocities close to the Sun. This can be illustrated with specific examples: for a stream diameter of \SI{20}{\degree}, the boundary region begins at about \SI{3}{\degree} from the center, so that peak velocities decrease beyond this latitude. For a larger stream with \SI{30}{\degree}, the boundary region starts at about \SI{8}{\degree}, and reduced velocities are only observed beyond this latitude. The key point is that lower observed peak velocities do not indicate weaker acceleration at the Sun, but instead reflect the increasing contribution of boundary-region plasma with latitude and interplanetary deceleration of the fastest plasma parcels during propagation.

The travel time of the high-speed streams from the Sun to \SI{1}{AU} is shown in panel~(b). It varies between $2.8$ and \SI{3.6}{days}, which falls within the commonly reported observational range of $2$ to $5$ days, but covers a narrower interval. This difference arises from the definition of travel time. The travel time of a high-speed stream is the travel time of its front, i.e., the time interval between the release of the first high-speed stream plasma parcel in a given radial direction and its arrival at the spacecraft. This parcel originates from the leading boundary of the stream close to the Sun and propagates with the velocity of the stream interface to the spacecraft. The parcel's velocity is therefore governed by the propagation speed of the stream interface, which varies only moderately (see Section~\ref{sec:si}). 

In contrast, observational studies typically define travel times as the interval between the time when the center of the source coronal hole passes the central meridian and the arrival of the peak velocity at Earth. This does not correspond to the travel time of the high-speed stream front. Therefore, such observational definitions capture timing differences between easily identifiable features at the Sun and at the spacecraft, rather than the physical propagation time of the stream front.

The minimum density at \SI{1}{AU} is shown in panel~(c). The global minimum density, measured in the stream center, is highest for small streams with diameters \SI{<14}{\degree} and decreases with increasing diameter until it converges to about \SI{0.7}{cm^{-3}}. The local minimum density, as measured by a virtual spacecraft, is lowest in the stream center and increases toward the flanks.

If only spherical expansion were present, the minimum density at \SI{1}{AU} in our setup would be \SI{1.5}{cm^{-3}}. However, the simulations show values between $0.7$ and \SI{3.7}{cm^{-3}}, which can be explained by two effects. First, for small streams or for measurements in the flanks, the spacecraft samples boundary-region plasma, which has enhanced density already close to the Sun (Section~\ref{sec:closeSun}). This leads to minimum densities above \SI{1.5}{cm^{-3}}. Second, densities below \SI{1.5}{cm^{-3}} indicate additional radial expansion beyond spherical expansion. These plasma parcels originate from the trailing boundary region close to the Sun, which evolves into the stream tail. There, velocity gradients lead to a radial dispersion of plasma parcels and thus density depletion, as has been described in Section~\ref{sec:vnt}.

The peak density in the stream interaction region is shown in panel~(d). It follows roughly the distribution of the peak velocity. The global peak density, occuring in the center of the stream interaction region, increases approximately linearly with stream diameter for small streams \SI{<14}{\degree}; beyond this value, the increase slows down and eventually saturates for large streams. The local peak density, measured at a given latitude by the virtual spacecraft, depends on the latitudinal position within the stream: it is largest in the stream center and decreases toward the flanks.

The overall trend of the peak density follows that of the peak velocity because it is directly controlled by it. The peak density forms through compression of slow solar wind plasma ahead of the stream, with the degree of compression set by the peak velocity. In contrast, the latitudinal variation is weaker because the stream interface deflects plasma and redistributes it toward the flanks, thereby partially compensating the stronger center-to-flank contrast present in the peak velocity (Section~\ref{sec:si}).

The peak temperature is shown in panel~(e). The global peak temperature, occuring in the latitudinal center of the high-speed stream behind the stream interface, increases rapidly with stream diameter and subsequently saturates. The local peak temperature as measured by a satellite depends strongly on the spacecraft latitude and decreases fast from the stream center towards its flanks. This behavior reflects the thermalization of high-speed stream plasma in the interaction region: the available energy for heating scales with the kinetic energy of the plasma parcels, and thus with the square of their velocity. As a result, high velocities in the center of sufficiently large streams lead to strong heating, while lower velocities in small streams or in the flanks result in significantly lower temperatures.

The peak magnetic field strength in the stream interaction region is shown in panel~(f). Both the global and local distributions closely follow those of the peak density, as the magnetic field is frozen into the solar wind plasma. As a result, both quantities are controlled by the same compression processes, leading to their similar distributions.

Overall, these results demonstrate that the local properties measured by a spacecraft can differ substantially from the global properties of the high-speed stream, and that they depend strongly on both the stream diameter and the sampling location within the stream. This is particularly notable given that all streams were initialized with identical plasma conditions at the inner boundary, highlighting how stream geometry and sampling position controls the local plasma properties observed in situ.

\subsection{Geoeffectivity}

\begin{figure}[t!]
 \includegraphics[width=\textwidth]{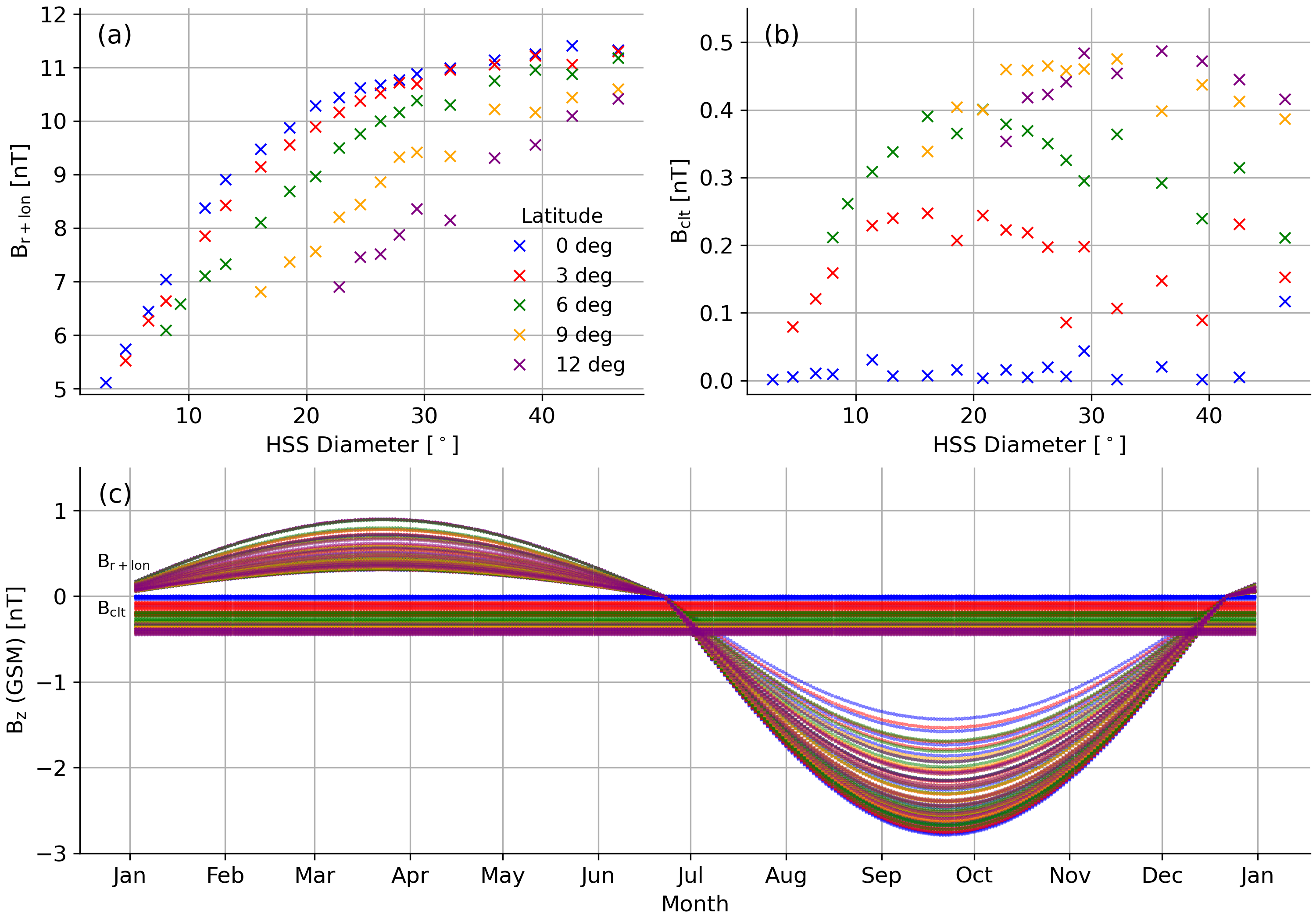}
\caption{Formation of the southward magnetic field component along Earth’s magnetic field axis. (a)~Peak heliospheric equatorial magnetic field strength and (b)~peak colatitudinal magnetic field strength of the high-speed stream as functions of stream diameter and latitudinal position within the stream. (c)~Projection of (a) and (b) onto Earth’s magnetic field axis, which varies with the season of the year.}
\label{fig:dst_B}
\end{figure}

\begin{figure}[t!]
 \includegraphics[width=\textwidth]{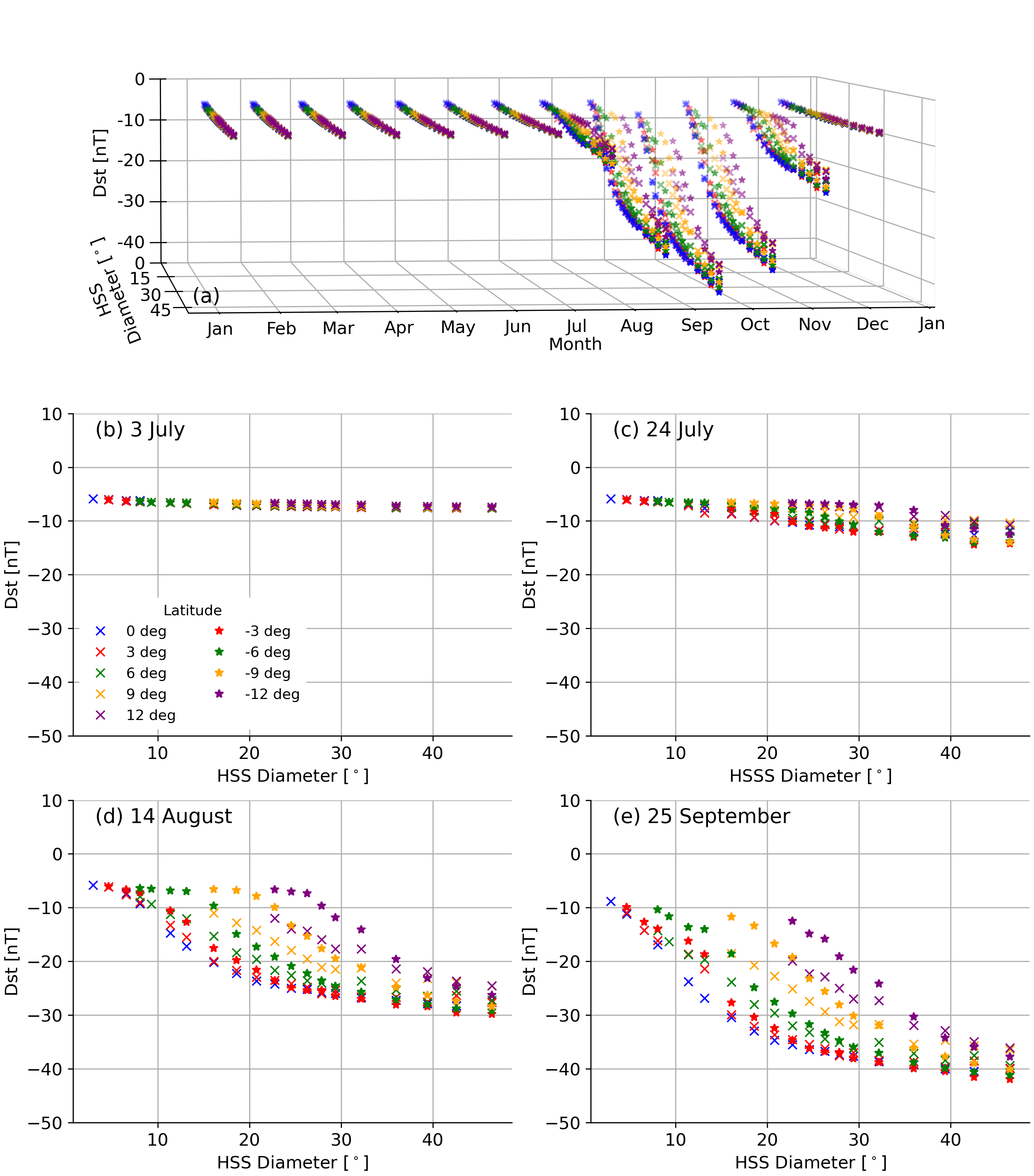}
\caption{(a)~Estimated Dst index as a function of stream diameter, latitudinal position within the stream, and season. Panels~(b)–(e) show daily slices of (a) for July~3, July~24, August~14, and September~25.}
\label{fig:Dst}
\end{figure}

We investigate the geoeffectivity of high-speed streams using the geomagnetic Dst index, which primarily measures the strength of Earth's ring current. The geoeffectivity of a high-speed stream is mainly controlled by its velocity and by the southward magnetic field component along Earth's magnetic field axis, as described in Section~\ref{sec:setup} by Equation~\ref{eq:dst}. In the following, we distinguish between the southward magnetic field component, defined along Earth's magnetic field axis, and the colatitudinal component, defined in the heliocentric coordinate system. These are not equivalent, since Earth's magnetic field axis is inclined relative to the ecliptic and changes its orientation over the year.

We first analyze the formation and strength of the southward magnetic field component as a function of stream diameter and position within the stream. The corresponding velocity dependence was discussed in Section~\ref{sec:1au}. The southward component arises from the projection of the solar wind magnetic field onto Earth's magnetic field axis. To better separate the underlying physical contributions, we decompose the solar wind magnetic field into a component in the heliospheric equatorial plane and a component in the latitudinal direction.

The peak value of the magnetic field component in the heliospheric equatorial plane occurs in the compressed stream interaction region and is shown in Figure~\ref{fig:dst_B}(a). It varies from about $5$ to \SI{11}{nT}, depending on stream diameter and sampling location within the stream. As this component is composed of the radial and longitudinal magnetic field components, which also dominate the total magnetic field strength, its distribution closely follows that of the peak total magnetic field strength in Figure~\ref{fig:hss_1au}.

The peak colatitudinal magnetic field component, shown in Figure~\ref{fig:dst_B}(b), forms through latitudinal deflection of the magnetic field frozen into the solar wind at the stream interface. Its magnitude is much smaller, ranging from $0$ to \SI{0.5}{nT}, but it is not negligible. The colatitudinal component vanishes near the stream center, where latitudinal deflection is minimal. It increases toward the flanks due to the increasing curvature of the stream interface, which leads to stronger deflection of both the solar wind plasma and the frozen-in magnetic field.

For fixed latitudinal positions, the dependence on stream diameter shows a non-monotonic behavior, with a maximum appearing in each of the curves. These maxima arise from two competing effects. For small streams, increasing stream diameter enhances compression and thus increases the magnetic field strength, leading up to the maximum. For larger streams, a fixed sampling latitude at increasing stream size corresponds to a position that moves progressively closer to the stream center relative to the stream size, where latitudinal deflection is weaker and thus the colatitudinal magnetic field strength decreases.

The contribution of both components to the southward magnetic field depends on their projection onto Earth's magnetic field axis, which varies over the year. This is shown in panel~(c). The southward projection of the equatorial component reaches its most negative values near the end of September for our positive-polarity stream (and near the end of March for negative-polarity streams, which are not discussed here), when Earth's magnetic field direction and the solar wind Parker spiral direction are best aligned. Its magnitude ranges from about $+1$ to \SI{-2.7}{nT}, depending on stream size, sampling location, and season.

In contrast, the southward projection of the colatitudinal component does not vary with season, as it is independent of the longitudinal orientation of Earth's magnetic field axis. Its magnitude ranges from $0$ to \SI{-0.4}{nT}, depending on stream diameter and position within the stream. Its sign depends on both the latitudinal position within the stream and the polarity of the high-speed stream: it is negative for a positive-polarity stream north of the stream center as shown here, but would reverse its polarity when sampling south of the center or when the stream polarity is reversed.

The total southward magnetic field is given by the superposition of the projected equatorial and colatitudinal components. As a result, the colatitudinal component can either enhance or reduce the total southward field, depending on stream polarity and sampling location, thereby modulating the geoeffectivity. This is illustrated in Figure~\ref{fig:Dst}. Panel~(a) shows the seasonal variation of the Dst response, with a maximum response in late September for our case. Panels~(b)--(e) show selected snapshots throughout the year. On July~3rd, the longitudinal component of Earth's magnetic field axis is nearly perpendicular to the Parker spiral direction, so that the equatorial solar wind magnetic field component does not project onto Earth's magnetic axis, resulting in a minimal geomagnetic response. By July~24th, a weak response appears, which increases through August~14th and reaches a maximum around September~25th as the alignment between Earth's magnetic field axis and the Parker spiral direction improves.

In addition to this seasonal variation, the Dst response increases with stream diameter and decreases with distance from the stream center. This reflects the increase in solar wind velocity and magnetic field compression with increasing stream size, and their gradual decrease from the center toward the flanks of the stream. Furthermore, for our high-speed stream, the response is generally stronger, i.e., the Dst is more negative, when Earth is located north of the stream center than when it is south. This asymmetry arises from the colatitudinal magnetic field component, which is generated by latitudinal deflection of the solar wind magnetic field at the stream interface.

Finally, we note that MHD simulations tend to underestimate both compression and magnetic field deflection in the stream interaction region. Stronger compression would enhance the southward magnetic field and thus increase the Dst response. Stronger deflection would amplify the colatitudinal field component. The sign of this contribution depends on both the sampling latitude within the stream and the stream polarity. Consequently, whether the Dst response is enhanced or reduced in the northern or southern hemisphere of the stream, respectively, is determined by the combined effect of the latitudinal position within the stream and the polarity of the stream.

\section{Summary} \label{sec:summary}

In this study, we have used three-dimensional MHD simulations to investigate the propagation of high-speed solar wind streams from their origin near the Sun to their impact at Earth. By tracking both global stream structure and individual plasma parcels, we have clarified to what extent local in-situ measurements can be used to explain the underlying physical evolution of the stream. The main conclusions are summarized as follows:
\begin{itemize}

\item[(1)] High-speed streams are not parcel-preserving structures.  
Common observational diagnostics such as peak velocity, minimum density, or peak temperature do not track fixed plasma parcels. Instead, these quantities evolve dynamically as different regions of the stream (core, boundary, and tail) continuously replace each other due to compression and expansion. As a result, feature-based radial trends can differ substantially from the true underlying plasma evolution, whereas using the plasma velocity as a reference provides a more robust framework for linking plasma parcels across heliocentric distances.

\item[(2)] Stream interaction regions dominate the evolution of high-speed streams.  
Compression at the stream interface leads to deceleration of fast plasma, acceleration of slow solar wind, and strong heating. This interaction progressively reduces the peak velocity of the stream and enhances density and temperature in the interaction region, while plasma outside the stream interaction region propagates nearly ballistically.

\item[(3)] Boundary-layer formation close to the Sun strongly biases observations.  
A stable boundary layer forms close to the Sun as a dynamical effect of stream propagation. This boundary layer can occupy a significant fraction of narrow streams. As a result, spacecraft frequently sample boundary plasma rather than the stream core, particularly for smaller streams. This leads to systematically reduced observed peak velocities that are unrelated to the intrinsic acceleration at the Sun.

\item[(4)] Three-dimensional transport critically reshapes stream structure.  
Latitudinal flows at the stream interface redistribute mass, momentum, and magnetic flux from the stream center toward the flanks. This reduces the center-to-flank contrasts expected from one-dimensional arguments and demonstrates that realistic HSS evolution requires fully three-dimensional modeling.

\item[(5)] Radial magnetic flux is conserved, but total field strength is not.  
The radial magnetic field component conserves magnetic flux with heliocentric distance when properly integrated. In contrast, the total magnetic field strength is not flux-conserving in a spherical sampling geometry due to the increasing contribution of non-radial field components that decay more slowly than expected from spherical expansion.

\item[(6)] High-speed stream properties vary strongly with sampling location and stream geometry.  
This variability arises even though all streams are initialized with identical plasma properties near the Sun. Large streams develop well-defined velocity plateaus and allow access to core plasma when sampled near their center, whereas smaller streams are dominated by boundary regions and their evolution during interplanetary propagation. As a result, the properties observed by a spacecraft depend strongly on its location within the stream, leading to substantial differences between local measurements and the global high-speed stream plasma properties.

\item[(7)] Geoeffectivity is controlled by stream velocity and compression, with additional north–south asymmetries due to latitudinal deflection.  
The Dst response increases with velocity and magnetic field compression in the stream interaction region, and therefore depends on stream size and Earth’s position within the stream. Seasonal modulation arises from the projection of the Parker spiral magnetic field onto Earth’s dipole axis. Latitudinal magnetic field deflection at the stream interface modifies the generated southward field, enhancing or reducing its strength depending on stream polarity and  whether Earth is located in the northern or southern hemisphere of the stream.
\end{itemize}

These results demonstrate that the large-scale propagation of high-speed solar wind streams cannot be fully described using one-dimensional interpretations. Instead, their evolution is governed by a coupled set of processes involving the formation of the boundary-layer formation close to the Sun, deceleration of the fastest plasma parcels by the stream interaction regions, and three-dimensional plasma transport along the stream interface in the stream interaction region.

An important implication is that many observationally derived scaling relations, such as radial trends of density and temperature in interplanetary space, do not trace the physical evolution of plasma parcels but instead reflect a moving mixture of different stream regions. This impedes inferences of the underlying plasma evolution and the derivation of physical scaling laws for solar wind propagation through interplanetary space.

The simulations further show that geometry plays a central role in determining observed stream properties. The relative contribution of core and boundary plasma depends strongly on stream width and spacecraft latitude, implying that observational samples are inherently biased unless the stream’s global structure is taken into account. In particular, small streams may appear systematically weaker simply because they are dominated by boundary layers rather than because of differences in solar acceleration.

Three-dimensional mass transport at the stream interface drives latitudinal flows away from the stream center. These flows provide a diagnostic of the spacecraft’s position within the stream, indicating whether the stream center passes north or south of the observer. Because high-speed streams propagate approximately radially from the Sun, this also constrains whether the source coronal hole is located in the northern or southern solar hemisphere, providing an independent check for linking in-situ measurements to their solar source regions.

Finally, the results highlight the importance of fully three-dimensional effects for space weather prediction. Magnetic field deflection at the stream interface introduces hemispheric asymmetries in the geoeffective southward field component. Together with compression in the interaction region, this likely enhances the variability of geomagnetic response to high-speed streams beyond what is captured in standard MHD models.

Three-dimensional mass transport at the stream interface drives latitudinal flows away from the stream center. These flows provide a diagnostic of the spacecraft’s position within the stream, indicating whether the stream center passes north or south of the observer. Because high-speed streams propagate approximately radially from the Sun, this also constrains whether the source coronal hole is located in the northern or southern solar hemisphere, providing an independent check for linking in-situ measurements to their solar source regions.

Finally, the results highlight the importance of fully three-dimensional effects for both heliospheric structure and space weather prediction. The latitudinal transport within stream interaction regions not only redistributes plasma and magnetic flux, but the latitudinal deflection at the stream interface also introduces an asymmetry in the geoeffective magnetic field component dependent on whether Earth is located in the northern or southern hemisphere of the high-speed stream. The compression and the latitudinal deflection are likely to be underestimated in standard MHD models, suggesting that the true variability of geomagnetic response to high-speed streams may be larger than predicted here.

Overall, the study provides a unified physical framework for interpreting high-speed stream observations and emphasizes that both local sampling geometry and global stream evolution must be considered when relating in-situ measurements to solar sources.

\acknowledgements 

\bibliographystyle{aa} 

\bibliography{bibliography}

@ARTICLE{vasyliunas2006,
       author = {{Vasyliunas}, Vytenis M.},
        title = "{Reinterpreting the Burton-McPherron-Russell equation for predicting Dst}",
      journal = {Journal of Geophysical Research (Space Physics)},
         year = 2006,
        month = jul,
       volume = {111},
       number = {A7},
          eid = {A07S04},
        pages = {A07S04},
          doi = {10.1029/2005JA011440},
       adsurl = {https://ui.adsabs.harvard.edu/abs/2006JGRA..111.7S04V}
}

@ARTICLE{pomoell2018,
       author = {{Pomoell}, Jens and {Poedts}, S.},
        title = "{EUHFORIA: European heliospheric forecasting information asset}",
      journal = {Journal of Space Weather and Space Climate},
         year = 2018,
        month = jun,
       volume = {8},
          eid = {A35},
        pages = {A35},
          doi = {10.1051/swsc/2018020},
       adsurl = {https://ui.adsabs.harvard.edu/abs/2018JSWSC...8A..35P}
}

@ARTICLE{cranmer2002,
       author = {{Cranmer}, Steven R.},
        title = "{Coronal Holes and the High-Speed Solar Wind}",
      journal = {\ssr},
         year = 2002,
        month = aug,
       volume = {101},
       number = {3},
        pages = {229-294},
          doi = {10.1023/A:1020840004535},
       adsurl = {https://ui.adsabs.harvard.edu/abs/2002SSRv..101..229C}
}

@ARTICLE{tsurutani2006,
       author = {{Tsurutani}, Bruce T. and {Gonzalez}, Walter D. and {Gonzalez}, Alicia L.~C. and {Guarnieri}, Fernando L. and {Gopalswamy}, Nat and {Grande}, Manuel and {Kamide}, Yohsuke and {Kasahara}, Yoshiya and {Lu}, Gang and {Mann}, Ian and {McPherron}, Robert and {Soraas}, Finn and {Vasyliunas}, Vytenis},
        title = "{Corotating solar wind streams and recurrent geomagnetic activity: A review}",
      journal = {Journal of Geophysical Research (Space Physics)},
         year = 2006,
        month = jul,
       volume = {111},
       number = {A7},
          eid = {A07S01},
        pages = {A07S01},
          doi = {10.1029/2005JA011273},
       adsurl = {https://ui.adsabs.harvard.edu/abs/2006JGRA..111.7S01T}
}

@ARTICLE{krieger1973,
       author = {{Krieger}, A.~S. and {Timothy}, A.~F. and {Roelof}, E.~C.},
        title = "{A Coronal Hole and Its Identification as the Source of a High Velocity Solar Wind Stream}",
      journal = {\solphys},
         year = 1973,
        month = apr,
       volume = {29},
       number = {2},
        pages = {505-525},
          doi = {10.1007/BF00150828},
       adsurl = {https://ui.adsabs.harvard.edu/abs/1973SoPh...29..505K}
}

@ARTICLE{wiegelmann2004,
       author = {{Wiegelmann}, T. and {Solanki}, S.~K.},
        title = "{Similarities and Differences between Coronal Holes and the Quiet Sun: Are Loop Statistics the Key?}",
      journal = {\solphys},
         year = 2004,
        month = dec,
       volume = {225},
       number = {2},
        pages = {227-247},
          doi = {10.1007/s11207-004-3747-2},
archivePrefix = {arXiv},
       eprint = {0802.0120},
 primaryClass = {astro-ph},
       adsurl = {https://ui.adsabs.harvard.edu/abs/2004SoPh..225..227W}
}

@ARTICLE{zirker1977,
       author = {{Zirker}, J.~B.},
        title = "{Coronal holes and high-speed wind streams.}",
      journal = {Reviews of Geophysics and Space Physics},
         year = 1977,
        month = aug,
       volume = {15},
        pages = {257-269},
          doi = {10.1029/RG015i003p00257},
       adsurl = {https://ui.adsabs.harvard.edu/abs/1977RvGSP..15..257Z}
}

@ARTICLE{altschuler1969,
       author = {{Altschuler}, Martin D. and {Newkirk}, Jr., Gordon},
        title = "{Magnetic Fields and the Structure of the Solar Corona. I: Methods of Calculating Coronal Fields}",
      journal = {\solphys},
         year = 1969,
        month = sep,
       volume = {9},
       number = {1},
        pages = {131-149},
          doi = {10.1007/BF00145734},
       adsurl = {https://ui.adsabs.harvard.edu/abs/1969SoPh....9..131A}
}

@ARTICLE{wiegelmann2017,
       author = {{Wiegelmann}, Thomas and {Petrie}, Gordon J.~D. and {Riley}, Pete},
        title = "{Coronal Magnetic Field Models}",
      journal = {\ssr},
         year = 2017,
        month = sep,
       volume = {210},
       number = {1-4},
        pages = {249-274},
          doi = {10.1007/s11214-015-0178-3},
       adsurl = {https://ui.adsabs.harvard.edu/abs/2017SSRv..210..249W}
}

@ARTICLE{altschuler1977,
       author = {{Altschuler}, M.~D. and {Levine}, R.~H. and {Stix}, M. and {Harvey}, J.},
        title = "{High resolution mapping of the magnetic field of the solar corona.}",
      journal = {\solphys},
         year = 1977,
        month = mar,
       volume = {51},
       number = {2},
        pages = {345-375},
          doi = {10.1007/BF00216372},
       adsurl = {https://ui.adsabs.harvard.edu/abs/1977SoPh...51..345A}
}

@ARTICLE{levine1977,
       author = {{Levine}, R.~H.},
        title = "{Evolution of open magnetic structures on the Sun: the Skylab period.}",
      journal = {\apj},
         year = 1977,
        month = nov,
       volume = {218},
        pages = {291-305},
          doi = {10.1086/155682},
       adsurl = {https://ui.adsabs.harvard.edu/abs/1977ApJ...218..291L}
}

@ARTICLE{parker1965,
       author = {{Parker}, E.~N.},
        title = "{Dynamical Theory of the Solar Wind}",
      journal = {\ssr},
         year = 1965,
        month = sep,
       volume = {4},
       number = {5-6},
        pages = {666-708},
          doi = {10.1007/BF00216273},
       adsurl = {https://ui.adsabs.harvard.edu/abs/1965SSRv....4..666P}
}

@ARTICLE{kasper2021,
       author = {{Kasper}, J.~C. and {Klein}, K.~G. and {Lichko}, E. and {Huang}, Jia and {Chen}, C.~H.~K. and {Badman}, S.~T. and {Bonnell}, J. and {Whittlesey}, P.~L. and {Livi}, R. and {Larson}, D. and {Pulupa}, M. and {Rahmati}, A. and {Stansby}, D. and {Korreck}, K.~E. and {Stevens}, M. and {Case}, A.~W. and {Bale}, S.~D. and {Maksimovic}, M. and {Moncuquet}, M. and {Goetz}, K. and {Halekas}, J.~S. and {Malaspina}, D. and {Raouafi}, Nour E. and {Szabo}, A. and {MacDowall}, R. and {Velli}, Marco and {Dudok de Wit}, Thierry and {Zank}, G.~P.},
        title = "{Parker Solar Probe Enters the Magnetically Dominated Solar Corona}",
      journal = {\prl},
         year = 2021,
        month = dec,
       volume = {127},
       number = {25},
          eid = {255101},
        pages = {255101},
          doi = {10.1103/PhysRevLett.127.255101},
       adsurl = {https://ui.adsabs.harvard.edu/abs/2021PhRvL.127y5101K}
}

@ARTICLE{weber1967,
       author = {{Weber}, Edmund J. and {Davis}, Jr., Leverett},
        title = "{The Angular Momentum of the Solar Wind}",
      journal = {\apj},
         year = 1967,
        month = apr,
       volume = {148},
        pages = {217-227},
          doi = {10.1086/149138},
       adsurl = {https://ui.adsabs.harvard.edu/abs/1967ApJ...148..217W}
}

@ARTICLE{parker1958,
       author = {{Parker}, E.~N.},
        title = "{Dynamics of the Interplanetary Gas and Magnetic Fields.}",
      journal = {\apj},
         year = 1958,
        month = nov,
       volume = {128},
        pages = {664},
          doi = {10.1086/146579},
       adsurl = {https://ui.adsabs.harvard.edu/abs/1958ApJ...128..664P}
}

@ARTICLE{gosling1978,
       author = {{Gosling}, J.~T. and {Asbridge}, J.~R. and {Bame}, S.~J. and {Feldman}, W.~C.},
        title = "{Solar wind stream interfaces}",
      journal = {\jgr},
         year = 1978,
        month = apr,
       volume = {83},
       number = {A4},
        pages = {1401-1412},
          doi = {10.1029/JA083iA04p01401},
       adsurl = {https://ui.adsabs.harvard.edu/abs/1978JGR....83.1401G}
}

@ARTICLE{gonzalez1987,
       author = {{Gonzalez}, Walter D. and {Tsurutani}, Bruce T.},
        title = "{Criteria of interplanetary parameters causing intense magnetic storms ( D$_{st}$ < -100 nT)}",
      journal = {\planss},
         year = 1987,
        month = sep,
       volume = {35},
       number = {9},
        pages = {1101-1109},
          doi = {10.1016/0032-0633(87)90015-8},
       adsurl = {https://ui.adsabs.harvard.edu/abs/1987P&SS...35.1101G}
}

@ARTICLE{gonzalez1994,
       author = {{Gonzalez}, W.~D. and {Joselyn}, J.~A. and {Kamide}, Y. and {Kroehl}, H.~W. and {Rostoker}, G. and {Tsurutani}, B.~T. and {Vasyliunas}, V.~M.},
        title = "{What is a geomagnetic storm?}",
      journal = {\jgr},
         year = 1994,
        month = apr,
       volume = {99},
       number = {A4},
        pages = {5771-5792},
          doi = {10.1029/93JA02867},
       adsurl = {https://ui.adsabs.harvard.edu/abs/1994JGR....99.5771G}
}

@ARTICLE{russell1973,
       author = {{Russell}, C.~T. and {McPherron}, R.~L.},
        title = "{Semiannual variation of geomagnetic activity}",
      journal = {\jgr},
         year = 1973,
        month = jan,
       volume = {78},
       number = {1},
        pages = {92},
          doi = {10.1029/JA078i001p00092},
       adsurl = {https://ui.adsabs.harvard.edu/abs/1973JGR....78...92R}
}

@ARTICLE{fox2016,
       author = {{Fox}, N.~J. and {Velli}, M.~C. and {Bale}, S.~D. and {Decker}, R. and {Driesman}, A. and {Howard}, R.~A. and {Kasper}, J.~C. and {Kinnison}, J. and {Kusterer}, M. and {Lario}, D. and {Lockwood}, M.~K. and {McComas}, D.~J. and {Raouafi}, N.~E. and {Szabo}, A.},
        title = "{The Solar Probe Plus Mission: Humanity's First Visit to Our Star}",
      journal = {\ssr},
         year = 2016,
        month = dec,
       volume = {204},
       number = {1-4},
        pages = {7-48},
          doi = {10.1007/s11214-015-0211-6},
       adsurl = {https://ui.adsabs.harvard.edu/abs/2016SSRv..204....7F}
}

@ARTICLE{muller2020,
       author = {{M{\"u}ller}, D. and {St. Cyr}, O.~C. and {Zouganelis}, I. and {Gilbert}, H.~R. and {Marsden}, R. and {Nieves-Chinchilla}, T. and {Antonucci}, E. and {Auch{\`e}re}, F. and {Berghmans}, D. and {Horbury}, T.~S. and {Howard}, R.~A. and {Krucker}, S. and {Maksimovic}, M. and {Owen}, C.~J. and {Rochus}, P. and {Rodriguez-Pacheco}, J. and {Romoli}, M. and {Solanki}, S.~K. and {Bruno}, R. and {Carlsson}, M. and {Fludra}, A. and {Harra}, L. and {Hassler}, D.~M. and {Livi}, S. and {Louarn}, P. and {Peter}, H. and {Sch{\"u}hle}, U. and {Teriaca}, L. and {del Toro Iniesta}, J.~C. and {Wimmer-Schweingruber}, R.~F. and {Marsch}, E. and {Velli}, M. and {De Groof}, A. and {Walsh}, A. and {Williams}, D.},
        title = "{The Solar Orbiter mission. Science overview}",
      journal = {\aap},
         year = 2020,
        month = oct,
       volume = {642},
          eid = {A1},
        pages = {A1},
          doi = {10.1051/0004-6361/202038467},
archivePrefix = {arXiv},
       eprint = {2009.00861},
 primaryClass = {astro-ph.SR},
       adsurl = {https://ui.adsabs.harvard.edu/abs/2020A&A...642A...1M}
}

@ARTICLE{allen2021,
       author = {{Allen}, R.~C. and {Ho}, G.~C. and {Mason}, G.~M. and {Li}, G. and {Jian}, L.~K. and {Vines}, S.~K. and {Schwadron}, N.~A. and {Joyce}, C.~J. and {Bale}, S.~D. and {Bonnell}, J.~W. and {Case}, A.~W. and {Christian}, E.~R. and {Cohen}, C.~M.~S. and {Desai}, M.~I. and {Filwett}, R. and {Goetz}, K. and {Harvey}, P.~R. and {Hill}, M.~E. and {Kasper}, J.~C. and {Korreck}, K.~E. and {Lario}, D. and {Larson}, D. and {Livi}, R. and {MacDowall}, R.~J. and {Malaspina}, D.~M. and {McComas}, D.~J. and {McNutt}, R. and {Mitchell}, D.~G. and {Paulson}, K.~W. and {Pulupa}, M. and {Raouafi}, N. and {Stevens}, M.~L. and {Whittlesey}, P.~L. and {Wiedenbeck}, M.},
        title = "{Radial Evolution of a CIR: Observations From a Nearly Radially Aligned Event Between Parker Solar Probe and STEREO A}",
      journal = {\grl},
         year = 2021,
        month = feb,
       volume = {48},
       number = {3},
          eid = {e91376},
        pages = {e91376},
          doi = {10.1029/2020GL091376},
       adsurl = {https://ui.adsabs.harvard.edu/abs/2021GeoRL..4891376A}
}

@ARTICLE{damicis2026,
       author = {{D'Amicis}, R. and {Sorriso-Valvo}, L. and {Benella}, S. and {Panasenco}, O. and {Perrone}, D. and {Sioulas}, N. and {Velli}, M. and {Dhamane}, O.~S. and {De Marco}, R. and {Bruno}, R. and {Telloni}, D. and {Fortunato}, V. and {Mele}, G. and {Monti}, F. and {Owen}, C. and {Louarn}, P. and {Livi}, S.},
        title = "{Evolution of Alfv{\'e}nic slow wind parcels coming from the same solar source: observations by Parker Solar Probe, Solar Orbiter, and Wind}",
      journal = {\aap},
         year = 2026,
        month = feb,
       volume = {706},
          eid = {A153},
        pages = {A153},
          doi = {10.1051/0004-6361/202556711},
       adsurl = {https://ui.adsabs.harvard.edu/abs/2026A&A...706A.153D}
}

@ARTICLE{perrone2022,
       author = {{Perrone}, D. and {Perri}, S. and {Bruno}, R. and {Stansby}, D. and {D'Amicis}, R. and {Jagarlamudi}, V.~K. and {Laker}, R. and {Toledo-Redondo}, S. and {Stawarz}, J.~E. and {Telloni}, D. and {De Marco}, R. and {Owen}, C.~J. and {Raines}, J.~M. and {Settino}, A. and {Lavraud}, B. and {Maksimovic}, M. and {Vaivads}, A. and {Phan}, T.~D. and {Fargette}, N. and {Louarn}, P. and {Zouganelis}, I.},
        title = "{Evolution of coronal hole solar wind in the inner heliosphere: Combined observations by Solar Orbiter and Parker Solar Probe}",
      journal = {\aap},
         year = 2022,
        month = dec,
       volume = {668},
          eid = {A189},
        pages = {A189},
          doi = {10.1051/0004-6361/202243989},
       adsurl = {https://ui.adsabs.harvard.edu/abs/2022A&A...668A.189P}
}

@ARTICLE{perrone2019,
       author = {{Perrone}, Denise and {Stansby}, D. and {Horbury}, T.~S. and {Matteini}, L.},
        title = "{Radial evolution of the solar wind in pure high-speed streams: HELIOS revised observations}",
      journal = {\mnras},
         year = 2019,
        month = mar,
       volume = {483},
       number = {3},
        pages = {3730-3737},
          doi = {10.1093/mnras/sty3348},
archivePrefix = {arXiv},
       eprint = {1810.04014},
 primaryClass = {physics.space-ph},
       adsurl = {https://ui.adsabs.harvard.edu/abs/2019MNRAS.483.3730P}
}

@ARTICLE{henderson2025,
       author = {{Henderson}, Sarah A. and {Filwett}, Rachael J. and {Lee}, Christina O. and {Jolitz}, Rebecca and {Allen}, Robert C. and {Dayeh}, Maher A. and {Rahmati}, Ali and {Larson}, Davin and {Halekas}, Jasper S. and {Gruesbeck}, Jacob R. and {Galvin}, Antoinette and {Wei}, Hanying and {Heber}, Bernd},
        title = "{Examining the Radial Evolution of a Corotating Interaction Region Observed at STEREO-A and MAVEN}",
      journal = {\apj},
         year = 2025,
        month = oct,
       volume = {992},
       number = {1},
          eid = {34},
        pages = {34},
          doi = {10.3847/1538-4357/adfd55},
       adsurl = {https://ui.adsabs.harvard.edu/abs/2025ApJ...992...34H}
}

@ARTICLE{obrien2000,
       author = {{O'Brien}, T. Paul and {McPherron}, Robert L.},
        title = "{An empirical phase space analysis of ring current dynamics: Solar wind control of injection and decay}",
      journal = {\jgr},
         year = 2000,
        month = apr,
       volume = {105},
       number = {A4},
        pages = {7707-7720},
          doi = {10.1029/1998JA000437},
       adsurl = {https://ui.adsabs.harvard.edu/abs/2000JGR...105.7707O}
}

@ARTICLE{richardson2018,
       author = {{Richardson}, Ian G.},
        title = "{Solar wind stream interaction regions throughout the heliosphere}",
      journal = {Living Reviews in Solar Physics},
         year = 2018,
        month = dec,
       volume = {15},
       number = {1},
          eid = {1},
        pages = {1},
          doi = {10.1007/s41116-017-0011-z},
       adsurl = {https://ui.adsabs.harvard.edu/abs/2018LRSP...15....1R}
}

@ARTICLE{frost2022,
       author = {{Frost}, Anna Marie and {Owens}, Mathew and {Macneil}, Allan and {Lockwood}, Mike},
        title = "{Estimating the Open Solar Flux from In-Situ Measurements}",
      journal = {\solphys},
         year = 2022,
        month = jul,
       volume = {297},
       number = {7},
          eid = {82},
        pages = {82},
          doi = {10.1007/s11207-022-02004-6},
       adsurl = {https://ui.adsabs.harvard.edu/abs/2022SoPh..297...82F}
}

@ARTICLE{hofmeister2022,
       author = {{Hofmeister}, Stefan J. and {Asvestari}, Eleanna and {Guo}, Jingnan and {Heidrich-Meisner}, Verena and {Heinemann}, Stephan G. and {Magdalenic}, Jasmina and {Poedts}, Stefaan and {Samara}, Evangelia and {Temmer}, Manuela and {Vennerstrom}, Susanne and {Veronig}, Astrid and {Vr{\v{s}}nak}, Bojan and {Wimmer-Schweingruber}, Robert},
        title = "{How the area of solar coronal holes affects the properties of high-speed solar wind streams near Earth: An analytical model}",
      journal = {\aap},
         year = 2022,
        month = mar,
       volume = {659},
          eid = {A190},
        pages = {A190},
          doi = {10.1051/0004-6361/202141919},
archivePrefix = {arXiv},
       eprint = {2203.15689},
 primaryClass = {astro-ph.SR},
       adsurl = {https://ui.adsabs.harvard.edu/abs/2022A&A...659A.190H}
}

\end{document}